\RequirePackage{fix-cm}

\documentclass[a4paper,reqno,12pt]{amsart}
\usepackage[text={5.9in,8.5in},centering,top=30mm]{geometry}
\usepackage{mathrsfs}

\DeclareSymbolFont{cmsymbols}{OMS}{cmsy}{m}{n}
\SetSymbolFont{cmsymbols}{bold}{OMS}{cmsy}{b}{n}
\DeclareSymbolFontAlphabet{\mtc}{cmsymbols}

\usepackage{graphicx}
\usepackage{mathptmx}      % use Times fonts if available on your TeX system

\usepackage{bm}

\usepackage{amsmath,amsthm,amssymb}

\usepackage{times}

\usepackage{enumerate}
\usepackage{tabularx}
\usepackage[mathscr]{euscript}
\usepackage{eufrak}

\usepackage{stmaryrd}

\usepackage{pgf}
\usepackage{tikz}
\usetikzlibrary{patterns}
\usepackage[colorlinks=true,
            linkcolor=blue,
            urlcolor=black,
            citecolor=blue]{hyperref}

\providecommand{\D}{\mathbb}

\newcommand{\ii}{\mathrm{i}}
\providecommand{\prob}[1]{\D{P}\left\{ #1 \right\}}
\newcommand{\dist}{\mathrm{dist}}
\newcommand{\Ord}[1]{\mathrm{O}\left(#1\right)}
\newcommand{\ord}[1]{\mathrm{o}\left(#1\right)}

\newcommand{\card}{\mathrm{card}}

\newcommand{\supp}{\mathrm{supp\,}}

\DeclareMathOperator*{\essup}{ess\,sup}

\newcommand{\one}{\mathbf{1}}

\def\mcA{\mtc{A}}

\def\mcC{\mtc{C}}

\def\mcG{\mtc{G}}

\def\mcR{\mtc{R}}
\def\mcS{\mtc{S}}

\def\mcX{\mtc{X}}

\def\mcZ{\mtc{Z}}

\newcommand{\hlam}{\widehat{\lam}}

\newcommand{\mub}{\overline{\mathfrak{m}}}

\newcommand{\sigmab}{\overline{\sigma}}

\definecolor{redd}{rgb}{0.95,0.2,0.2}
\definecolor{gris}{rgb}{0.9,0.9,0.9}
\definecolor{greenn}{rgb}{0.1,0.6,0.2}

\definecolor{cmgray}{rgb}{0.7,0.7,0.7}

\definecolor{cmblue}{rgb}{0.2,0.5,0.8}

\newcommand{\be}{\begin{equation}}
\newcommand{\ee}{\end{equation}}
\newcommand{\ba}{\begin{array}{l}}
\newcommand{\ea}{\end{array}}

\newcommand{\bal}{\begin{aligned}}
\newcommand{\eal}{\end{aligned}}

\newcommand{\baln}{\begin{align}}
\newcommand{\ealn}{\end{align}}

\newcommand{\ble}{\begin{lemma}}
\newcommand{\ele}{\end{lemma}}

\newcommand{\bco}{\begin{cor}}
\newcommand{\eco}{\end{cor}}

\newcommand{\bpr}{\begin{proposition}}
\newcommand{\epr}{\end{proposition}}

\newcommand{\bre}{\begin{remark}}
\newcommand{\ere}{\end{remark}}

\newcommand{\btm}{\begin{thm}}
\newcommand{\etm}{\end{thm}}

\newcommand{\bde}{\begin{definition}}
\newcommand{\ede}{\end{definition}}

\newcommand{\lcite}[2]{{\cite[#2]{#1}}}

\newcommand{\myset}[1]{\left\{ \, #1 \, \right\}}

\newcommand{\eu}{{\mathrm{e}}}

\newcommand{\ffi}{\varphi}

\newcommand{\all}{\forall\,}

\newcommand{\half}{\frac{1}{2}}

\newcommand{\quart}{\frac{1}{4}}

\newcommand{\eps}{\epsilon}

\newcommand{\ups}{\varkappa}

\newcommand{\gam}{\boldsymbol{\gamma}}

\DeclareSymbolFont{newfont}{OML}{cmm}{m}{it}% Computer Modern math font
\DeclareMathSymbol{\Epsilon}{3}{newfont}{15}% Symbol 15
\DeclareMathSymbol{\Varrho}{3}{newfont}{37}% Symbol 37
\DeclareMathSymbol{\rro}{3}{newfont}{37}% Symbol 37

\newcommand{\tH}{\widetilde{H}}

\newcommand{\pt}{\partial}
\newcommand{\Const}{\mathrm{Const\,}}
\newcommand{\const}{\mathrm{const\,}}

\newcommand{\pr}[1]{\mathbb{P}\left\{ #1 \right\}}

\newcommand{\prsub}[2]{\mathbb{P}_{#1}\left\{ #2 \right\}}

\newcommand{\Bigpr}[1]{\mathbb{P}\Big\{ #1 \Big\}}

\newcommand{\esm}[1]{\D{E}\left[\, #1\, \right]}

\newcommand{\vertii}[1]{{\big\vert\kern-0.25ex\big\vert #1
    \big\vert\kern-0.25ex\big\vert\kern-0.25ex}}

\newcommand{\vertiii}[1]{{\big\vert\kern-0.25ex\big\vert\kern-0.25ex\big\vert #1
    \big\vert\kern-0.25ex\big\vert\kern-0.25ex\big\vert}}

\newcommand{\lam}{\lambda}
\newcommand{\om}{\omega}

\newcommand{\Bfq}{\pmb{\mathfrak{q}}}
\newcommand{\BfQ}{\pmb{\mathfrak{Q}}}

\newcommand{\Lam}{\Lambda}
\newcommand{\Om}{\Omega}

\newcommand{\hchi}{{\widehat{\chi}}}

\newcommand{\tB}{\widetilde{B}}

\newcommand{\lr}[1]{\langle#1\rangle}

\newcommand{\ball}{\mathrm{B}}

\newcommand{\fa}{\mathfrak{a}}

\newcommand{\fc}{\mathfrak{c}}

\newcommand{\fm}{\mathfrak{m}}

\newcommand{\fr}{\mathfrak{r}}
\newcommand{\fq}{\mathfrak{q}}

\newcommand{\fB}{\mathfrak{B}}

\newcommand{\fS}{\mathfrak{S}}
\newcommand{\fF}{\mathfrak{F}}

\newcommand{\fu}{\mathfrak{u}}

\newcommand{\heps}{{\widehat{\epsilon}}}

\newcommand{\BF}{\mathbf{F}}

\newcommand{\BU}{\mathbf{U}}

\newcommand{\cscB}{\mathcal{B}}

\newcommand{\csS}{\mathscr{S}}

\newcommand{\cA}{\mathcal{A}}

\newcommand{\cT}{\mathcal{T}}

\newcommand{\Br}{\mathbf{r}}

\newcommand{\ra}{\mathrm{a}}
\newcommand{\rc}{\mathrm{c}}

\newcommand{\rh}{\mathrm{h}}

\newcommand{\rrq}{\mathrm{q}}

\newcommand{\rL}{\mathrm{L}}

\newcommand{\rP}{\mathrm{P}}

\newcommand{\DC}{\mathbb{C}}
\newcommand{\DP}{\mathbb{P}}
\newcommand{\DR}{\mathbb{R}}

\newcommand{\DZ}{\mathbb{Z}}

\newcommand{\DN}{\mathbb{N}}

\newcommand{\etal}{\emph{et al.}\xspace}

\newcommand{\cond}{\,\big|\,}

\newcommand{\tto}[1]{\smash{\mathop{\,\,\,\, \longrightarrow \,\,\,\, }\limits_{#1}}}

\newenvironment{hyp}[1]{
\vskip3mm\par\noindent
$\blacklozenge$\;\textbf{Hypothesis #1.}
\noindent}

\newcommand{\bhy}[1]{\begin{hyp}{#1}}
\newcommand{\ehy}{\end{hyp}}

\newtheorem{thm}{Theorem}

\newtheorem{cor}{Corollary}%[section]

\newtheorem{proposition}[thm]{Proposition}

\usepackage{amssymb}
\usepackage{amsmath}
\usepackage{mathrsfs}

\usepackage{xspace}

\usepackage{epsfig}
\usepackage{enumerate}
\usepackage{graphicx}
\usepackage[title]{appendix}

\numberwithin{equation}{section}

\newtheorem{lemma}{Lemma}[section]

\newtheorem{remark}{Remark}[section]

\newtheorem{definition}{Definition}[section]

\begin{document}

\title[Weakly screened models]
{Density of States under non-local interactions II.
\\Simplified polynomially screened interactions}
%\subtitle{Local interpolation}
%Universality of smoothness of Density of States in arbitrary higher-dimensional disorder under non-local interactions I. From Vi\'{e}te--Euler identity to Anderson localization

%%%\titlerunning{Universal continuity properties of the DS under long-range interactions}        % if too long for running head

%%\author{Victor Chulaevsky\\\\Universit\'{e} de Reims\\
%%D\'{e}partement de math\'{e}matiques\\
%%51687 Reims Cedex, France}

\author{Victor Chulaevsky}

\address{Universit\'{e} de Reims\\
D\'{e}partement de math\'{e}matiques\\
51687 Reims Cedex, France}

\email{victor.tchoulaevski@univ-reims.fr}


%%%\institute{V. Chulaevsky \at
%%%              Universit\'{e} de Reims, D\'{e}partement de math\'{e}matiques \\
%%%              Moulin de la Housse, B.P. 1039,
%%%              51687 Reims Cedex, France\\
%%%%
%%%%%              Tel.: +123-45-678910\\
%%%%%              Fax: +123-45-678910\\
%%%              \email{victor.tchoulaevski@univ-reims.fr}           %  \\
%%%%             \emph{Present address:} of F. Author  %  if needed
%%%%%           \and
%%%%%           S. Author \at
%%%%%              second address
%%%}
%%\affil{Universit\'{e} de Reims\\D\'{e}partement de math\'{e}matiques\\
%%              51687 Reims Cedex, France}

\date{\today}

\maketitle

\begin{abstract}
Following [5], we analyze regularity properties of single-site probability distributions of the
random potential and of the
Integrated Density of States (IDS) in the Anderson models with infinite-range interactions.
In the present work, we study in detail a class of polynomially decaying interaction potentials
of rather artificial (piecewise-constant) form, and give a complete proof of infinite smoothness
of the IDS in an arbitrarily large finite domain subject to the fluctuations of the entire,
infinite random environment. A variant of this result, based as in [5] on the harmonic analysis
of probability measures, results in a proof of spectral and dynamical Anderson localization in the
considered models.

%%Insert your abstract here. Include keywords, PACS and mathematical
%%subject classification numbers as needed.
%%%\keywords{Density of states\and long-range interactions \and Coulomb screening \and  Anderson localization}
%%%\subclass{46N50 \and , 60B20}
\end{abstract}

%\tableofcontents

\section{Introduction}
\label{sec:intro}

This text is a follow-up of \cite{C16e}, where the reader can find the main motivations,
a historical review, and a number of bibliographical references.

The main model is an Anderson Hamiltonian with an infinite-range alloy-type random potential.
This corresponds to the physical reality where the fundamental interactions are most certainly \emph{not} compactly supported. By interactions we mean here those between the quantum particle (e.g., an electron),
evolving in a sample of a disordered media, with surrounding ions. Physically speaking, other mobile electrons
present in the media also make a non-negligible contribution to the potential field affecting the main particle
under consideration, especially from the point of view of the screening phenomena, substantially attenuating
the "naked" Coulomb potentials, but the full-fledged many-body spectral problem is beyond the scope of the
present paper. In this particular physical metaphor, the interaction potential means the screened Coulomb
potential. As discussed in
\cite{C16e}, the rate of its decay is not universal and varies from one physical model to another.

A large class of interaction potentials was considered in \cite{C16e}, but in the present, relatively short paper,
we focus on the particular case of interaction potentials featuring a summable \emph{power-law}
decay, and provide all technical details that were missing or only briefly outlined
in a general discussion of long-range Anderson models in \cite{C16e}.

Let be given a function
$$
x \mapsto \sum_{y \in \DZ^d} \fq_y \fu(x-y),
$$
which we will always assume absolutely summable on $\DZ^d$; more precisely, we assume that
$0 \le \fu(r) \le C r^{-A}$ for some $ A>d$. Then one can define
a linear transformation $\BU$, well-defined on any bounded function $\Bfq:\, \DZ^d \to \DR$:
\be
\label{eq:def.BU}
\BU:\, \Bfq \mapsto \BU[\Bfq] = V\,, \;\; V:\,\DZ^d \to \DR \,,
\ee
where
\be\label{eq:def.BU.2}
V(x) = (\BU[\Bfq])(x) = \sum_{y\in\DZ^d} \fu(y-x) \fq_y .
\ee

To clarify the main ideas of \cite{C16a} and simplify some technical aspects,
the interaction potential $\fu:\, \DR_+ \to \DR$ is assumed to have the following form.
Introduce a growing integer sequence $\fr_k = \lfloor k^\ups \rfloor$, $\ups>1$, $k\ge 0$, and let
\be
\label{eq:def.fu}
\fu(r) = \sum_{k=1}^{\infty} \fr_k^{-A} \one_{[\fr_{k}, \fr_{k+1})}(r)  \,,
\ee
Making $\fu(\cdot)$ piecewise constant will allow us to achieve,
albeit in a somewhat artificial setting, an elementary derivation of infinite smoothness of the DoS
from a similar property of single-site probability distributions of the potential $V$. We refer to $V$
as the \emph{cumulative} potential in order to distinguish it from the \emph{interaction} potential $\fu$
(which is a functional characteristics of the model) and from the local potential \emph{amplitudes}
$\{\fq_y, \, y\in\DZ^d \}$. The notation $\fq_y$ will be used in formulae and arguments
pertaining to general functional aspects of the model, while in the situation where the latter amplitudes are random
we denote them by $\om_y$.

We always assume the amplitudes $\fq_y$ and $\om_y$ to be uniformly bounded. In the case of random amplitudes,
one should either to assume this a.s. (almost surely, i.e., with probability one) or to construct from the beginning
a product measure on $[0, 1]^{\DZ^d}$ rather than on $\DR^{\DZ^d}$ and work with samples $\bm{\om}\in [0, 1]^{\DZ^d}$,
which are thus automatically bounded. It is worth mentioning that boundedness is not crucial to most of the key
properties established here, but results in a streamlined and more transparent presentation. On the other hand,
as pointed out in \cite{C16e}, there are interesting models with unbounded  amplitudes $\om_\bullet$
such that $\esm{ \left(\om_\bullet - \esm{\om_\bullet}\right)^2} <\infty$. It is readily seen that
single-site probability distribution of the cumulative potential $V(x,\om)$, necessarily compactly supported
when $\om_\bullet$ are uniformly bounded and the series \eqref{eq:def.BU.2} (with $\fq_y$ replaced with
$\om_y$) converges absolutely, cannot have an analytic density, for it would be compactly supported and not
identically zero, which is impossible. However, in some class of marginal measures of $\om_\bullet$
with unbounded support, considered long ago by Wintner \cite{Wint1934} in the framework of Fourier analysis of
probability measures, the single-site density of $V(\cdot,\om)$ can be analytic on $\DR$.

\vskip2mm
We also always assume that  $\om_y$ are IID. Extensions
to dependent random fields $(\om_x)_{x\in\DZ^d}$ with rapidly decaying correlations
do not really pose any serious problem, as limit theorems for normalized sums of random variables (r.v.)
are well-known to hold true for a large class of dependent random fields. Following the program outlined in \cite{C16e},
we plan to address such models in a separate work,
in a general context of Gibbs measures on the samples $\bm{\om}$.

\section{Main results}
\label{sec:main.results}

\subsection{Infinite smoothness of single-site distributions}

\btm
\label{thm:infinite.smooth.V}
Consider the potential $\fu(r)$ of the form \eqref{eq:def.fu}, with $A>d$ and let $d \ge 1$.
Then the characteristic functions of the random variables $V(x,\om)$ of the form \eqref{eq:def.BU.2}
obey the upper bound
$$
\big| \ffi_{V_x}(t)\big| \le \Const \eu^{ - c |t|^{d/A}} .
$$
Consequently, for any $d \ge 1$ the r.v. $V_x$ have probability  densities
$\rho_x \in \mcC^\infty(\DR)$.
\etm

\subsection{Infinite smoothness of the DoS}

\btm
\label{thm:DoS.xi.Lam}
Fix a bounded connected subset $\Lam \subset \DZ^d$.
\begin{enumerate}[\rm(A)]
  \item There exists a $\sigma$-algebra $\fB_\Lam$,
  an $\fB_\Lam$-measurable self-adjoint random operator $\tH_\Lam(\om)$ acting in $\ell^2(\Lam)$, and a
  $\fB_\Lam$-independent real-valued r.v. $\xi_\Lam$ such that
\be
\label{eq:H.H0.xi}
H_\Lam(\om) = \tH_\Lam(\om) + \xi_\Lam(\om) \one_\Lam \,.
\ee

  \item The characteristic function $\ffi_{\xi_\Lam}$ of \,$\xi_\Lam$ \, fulfills the decay bound
\be
 \big|  \ffi_{\xi_\Lam}(t) \big| \le C\, \eu^{-|t|^{d/A}} \,.
\ee
\end{enumerate}

\etm

\subsection{Wegner estimate}

\btm["Frozen bath" Wegner estimates]
\label{thm:Wegner}
Fix real numbers $\tau>1$ and $\theta\in(0, \tau-1)$,
consider a ball $\ball=\ball_L(u)$, and let
\be
\label{eq:thm.Wegner.eps}
R_L = R_L(\tau) = L^{\tau} , \quad
\eps_L = R_L^{- \frac{A}{1+\theta}} \,;
\ee
here  $\theta>0$ can be chosen arbitrarily small, for $L$ large enough.
Next, consider the Hamiltonian $H_\ball$ and a larger set
$\mcA = \mcA_L(\tau) = \ball_{R_L}(u)\setminus\ball_{L}(u)$,
and introduce the the product probability space $(\Om_\mcA, \fF_\mcA, \DP_\mcA)$
generated by the r.v. $\om_y$ with $y\in\mcA$.
Then
\be
\label{eq:cor.Wegner.bound}
\prsub{\mcA}{ \dist(\Sigma_{\ball}, E) \le \eps_L } \le C \, | \ball | \, \eps_L
\le C\, |\ball|\, \eps^\beta \,,
\ee
with
\be
\label{eq:beta.theta.tau}
\beta = 1 - \frac{1+\theta}{\tau} \in(0,1).
\ee
\etm

\bre
\label{rem:beta}
The constant $C$ in \eqref{eq:cor.Wegner.bound} can be absorbed in the exponent $\beta$ (in \eqref{eq:beta.theta.tau}),
by taking a slightly smaller $\theta>0$ and letting $L$ be large enough.
\ere

\subsection{Localization}

Below we denote by $\cscB_1(\DR)$ the set of all bounded Borel functions $\phi:\, \DR\to\DC$
with $\| \phi \|_\infty\le 1$.
\btm
\label{thm:main.polynom.large.g}
Consider the potential $\fu(r)$ of the form \eqref{eq:def.fu}, with $A>d$, and let $d>1$.
For any $m>0$ there exist $L_*\in\DN$ and $g_0>0$ such that for all $g$ with $|g|\ge g_0$
with probability one, the random operator
$H(\om)=-\Delta + gV(x,\om)$ has pure point spectrum with exponentially decaying eigenfunctions,
and for any $x,y\in\DZ^d$ and any connected subgraph $\mcG \subseteq \DZ^d$ containing $x$ and $y$
one has
\be
\esm{ \sup_{\phi \in\cscB_1(\DR)}
\; \big\| \one_x \phi\big(H_\mcG(\om)\big) \one_y  \big\| } \le \frac{C'}{ (1+|x-y|)^{C}} .
\ee
\etm

\btm
\label{thm:main.polynom.low.E}
Consider the potential $\fu(r)$ of the form \eqref{eq:def.fu}, with $A>d$, and let $d>1$.
There exist an energy interval $I = [E_0, E_0 + \eta]$, $\eta>0$, near the a.s. lower edge of spectrum
$E_0$ of the random operator $H(\om) = -\Delta + V(x,\om)$
such that
with probability one, $H(\om)$ has in $I$ pure point spectrum with exponentially decaying eigenfunctions,
and for any $x,y\in\DZ^d$ and any connected subgraph $\mcG \subseteq \DZ^d$  containing $x$ and $y$ one has
\be
\esm{ \sup_{\phi \in\cscB_1(\DR)}
\left\| \one_x \, \rP_I\big(H_\mcG(\om)\big) \phi\big(H_\mcG(\om)\big)  \,\one_y  \right\| }
   \le \frac{C'}{ (1+|x-y|)^{C}} .
\ee
\etm

\section{Fourier analysis}
\label{sec:char.f.polynom}

\subsection{The Main Lemma}

\ble
\label{lem:Main}
Let be given a family of IID r.v.
$$
X_{n,k}(\om), \; n\in \DN, \;\; 1 \le k \le K_n \,, \;\; K_n \asymp n^{d-1},
$$
and assume that their common characteristic function $\ffi_X(t) = \esm{ \eu^{\ii t X}}$ fulfills
\be
\label{eq:Main.Lemma.ffi.t0}
\ln \, \big| \ffi_X(t) \big|^{-1} \ge C_X t^2, \;\; |t|\le t_0.
\ee
Let
$$
\bal
S(\om) &= \sum_{n\ge 1} \sum_{k=1}^{K_n} \fa_n X_{n,k}(\om), \;\; \fa_n \asymp n^{-A} \,,
\\
S_{M,N}(\om) &= \sum_{n=M}^N \sum_{k=1}^{K_n} \fa_n X_{n,k}(\om), \;\; M \le N \,.
\eal
$$
The the following holds true.
\begin{enumerate}[\rm(A)]
  \item\label{item:Main.Lemma.A}
There exists $C = C(C_X, t_0, A,d)\in(0,+\infty)$ such that
$$
\all t\in\DR  \quad \big| \ffi_S(t) \big| \le C \eu^{- |t|^{d/A}} \,.
$$

  \item\label{item:Main.Lemma.B}
   For any $\eps>$, $N \ge (1+c)M \ge 1$ with $c>0$,  and $t$ with $|t| \le N^{A - \eps}$,
$$
S_{M, N} := \ln \left| \esm{ \eu^{\ii t S_{M,N}(\om)}} \right|^{-1}
\ge C  M^{-2A+d} \,t^2 \,.
$$

  \item\label{item:Main.Lemma.C}
  Let $I\subset \DR$ be an interval of finite length $|I|$. Then
  for any r.v. $Y$ independent of $S_{N,2N}$, one has
\be
\label{eq:Main.Lemma.D}
|I| \ge N^{- \frac{A}{1+\theta}} \quad \Longrightarrow \quad
\pr{ Y(\om) + S_{M,N}(\om) \in I } \le C \, M^A \, |I| \,,
\ee
where $\theta>0$ can be chosen arbitrarily small, provided $N$ is large enough: $N \ge N_*(\theta)$.
\end{enumerate}
\ele

\proof
By the IID property of the family $\{ \om_x, \, x\in\mcZ\}$, we have
$$
\bal
\ffi_{S}(t) &= \esm{ \exp\left( \ii t \sum_{n\ge 1} \sum_{k=1}^{K_n} \fa_n X_{n,k}(\om) \right)}
%%\\
%%%
%%&
= \prod_{n\ge 1} \prod_{k=1}^{K_n} \esm{ \eu^{ \ii t \fa_n X_{n,k}}}
\\
&
= \prod_{n\ge 1} \prod_{k=1}^{K_n} \ffi_ X\big( t \fa_{n}\big)
= \prod_{n\ge 1}  \left(\ffi_ X\big( t \fa_{n}\big)\right)^{K_n} \,,
\eal
$$
so for the logarithm we have the lower bound
\be
\label{eq:Main.Lemma.ln.ffi.b.1}
\bal
\ln \big| \ffi_S(t) \big|^{-1} &= \sum_{n\ge 1} K_n  \ln \big| \ffi_X\big( \fa_{n} t \big) \big|^{-1}
%%\\
%%%
%%&
\ge C_1 \sum_{n\ge 1} n^{d-1}  \ln \big| \ffi_X\big( \fa_{n} t \big) \big|^{-1}
\\
&
\ge C_1 \left(\sum_{n=1} + \sum_{n>N_t} \right) n^{d-1}  \ln \big| \ffi_X\big( \fa_{n} t \big) \big|^{-1}
=: \mcS_1(t) + \mcS_2(t) \,,
\eal
\ee
where all terms in $\mcS_1$ and in $\mcS_2$ are non-negative, since $|\ffi_X(t)|\le 1$ for any $t$.
In particular, this implies that $\ln \big| \ffi_S(t)\big|^{-1} \ge \min\big[ \mcS_1(t), \, \mcS_2(t) \big]$.
Now focus on $\mcS_2(t)$ and recall that, by definition of the threshold $N_t$ (cf. assertion (B)),
$$
\all n \ge N_t \quad n^{-A} |t| \le N_t^{-A} |t| \in[0, t_0] \,,
$$
hence, by hypothesis \eqref{eq:Main.Lemma.ffi.t0},
$$
\bal
\mcS_2(t) =
C_1 \sum_{n > N_t} \ln \big| \ffi_X\big( \fa_{n} t \big) \big|^{-1}
&
\ge C_2 \, t^2 \,  \sum_{n > N_t}  n^{d-1} \fa^2_{n}
\ge C_3 \, t^2 \,  \sum_{n > N_t}  n^{-2A+d-1}
\\
&
\ge C_4 t^2 \, N_t^{-2A+d}
\ge C_5 t^2 \, |t|^{-\frac{2A - d}{A}}
\\
&
=C_5 \, |t|^{d/A} \,,
\eal
$$
which proves assertion \eqref{item:Main.Lemma.A}.

In Section \ref{ssec:Wintner} we comment on Wintner's approach \cite{Wint1934}
to the estimation of $\mcS_1(t)$, based on an elementary lemma by P\'olya and Szeg\"{o}
\cite{PolSzego25}. The final result is similar : $\mcS_1(t) \ge C |t|^{d/A}$.

Assertion \eqref{item:Main.Lemma.B} is obtained in the same way:
$$
\bal
\sum_{n=M}^{N} \ln \big| \ffi_X(\fa_n t)\big|^{-1}
&
\ge C_1 t^2 \sum_{n=M}^{N} n^{-2A+d-1}
\ge C_2 t^2  \int_{n=N/(1+c)}^{N} s^{-2A+d-1} \, ds
\\
&
\ge C_3(A,d,c) \, t^2 \, N^{-2A+d} \ge C_4 |t|^{d/A} \,.
\eal
$$
Observe that for $N=M$, or close to $M$, we would have a weaker lower bound by
$C' |t|^{\frac{d-1}{A}}$, but with $d>1$ this is still good enough for the proof of infinite
derivability. This also works when $d>1$ is non-integer and arbitrarily close to $1$. One possible setting
where this observation can be useful is
a subset of $\DZ^d$ with the rate of growth of balls $r\mapsto r^{1+\delta}$, $\delta>0$.

%%Assertion (C) is essentially a reformulation of (B), where the roles of principal and subordinate parameters
%%are changed, but the main relation between the magnitude of $|t|$ and the indices $n$ in the
%%terms forming the key sum $\mcS_2(t)$ remains the same as in the above proofs of (A) and (B):
%%$$
%%|t| < N^{A-\eps} \ll N \quad  \Longrightarrow \quad
%%\all n\in[N, 2N] \quad \ln \big| \ffi_X(\fa_n t)\big|^{-1} \ge C \, t^2 \fa_n^2\,.
%%$$

Now we turn to the proof of assertion \eqref{item:Main.Lemma.C}. It suffices to consider the case where $Y(\om)=\const$ a.s.,
otherwise one can first condition on $Y$ (independent of $S_{M,N}$ by hypothesis), thus rendering $Y$
constant.
We need to assess the integrals of the probability measure of $S(\om)$ on intervals $I_\eps$ of length
$\Ord{\eps}$. It will be clear from the calculations given below that it sufficed to consider the case where
$I_\eps$ is centered at origin; a shift results in factors of unit modulus, so we stick to
$I_{[-\eps,\eps]}$ to have less cumbersome formulae. Further, since the main estimate will be achieved
in the Fourier representation, it is more convenient to work with a smoothed indicator function
$\chi_\eps$:
$$
\bal
\one_{I_\eps} & \le  \one_{[-2\eps, 2\eps]} \le \one_{[-4\eps, 4\eps]} \, \frac{1}{2\eps} \one_{[-\eps, \eps]}
\\
&
\le \chi_\eps := \one_{[-4\eps, 4\eps]}  \cdot \frac{\eu^{- \sigma_\eps^2 t^2/2}}{\sqrt{2\pi} \sigma_\eps} \,,
\eal
$$
with $\sigma_\eps = a\eps$, $a \approx 1.2$. The last inequality is easily validated by an elementary
numerical calculation: due to monotone decay of the Gaussian density on the positive half-axis, it suffices
to check its lower bound by $1$ on $[0,4\eps]$. The aim is of course
to secure a much faster decay at infinity for the Fourier transform than in the case of a discontinuous
indicator function. By the Parseval identity,
for any $\cT_\eps>0$,
$$
\bal
\mu_S(I_\eps) &= \int_\DR \one_{I_\eps}(x) \, dF_S(x)\le \int_\DR \chi_\eps(x) \, dF_S(x)
\\
&
\le  \int_{ |t|\le \cT_\eps}  \big| \hchi_\eps(t) \, \ffi_S(t) \big| \, dt
+ \int_{ |t| > \cT_\eps}  \big| \hchi_\eps(t) \, \ffi_S(t) \big| \, dt  =: J_1 + J_2 \,,
\eal
$$
where
$$
\hchi_\eps(t) = \eps \frac{ \sin(4\eps t)}{\eps t} \eu^{ - \frac{\sigma_\eps^2 t^2}{2}} \,.
$$
Now define $T_N$ by
\be
\label{eq:def.T.N}
T_N = \inf\myset{ t>0:\, \fa_N t \le t_0 }
= \inf\myset{ t>0:\, \lfloor N^{-A} \rfloor t \le t_0 }
\sim (C + \ord{1}) N^A \,,
\ee
(here $\ord{1}$ refers to $N\to \infty$) with some $C = C(t_0)\in(0,+\infty)$.

\vskip1mm
\noindent
$\bullet$ \textbf{Bound on $J_2$.}
Further, assume that $\eps>0$ is such that
\be
\label{eq:cT.eps}
\cT_\eps :=\eps^{-1} \ln^2 \eps^{-1} \le T_N \,,
\ee
(we shall see in a moment that it means $\eps$ is not too small),
then
\be
\label{eq:bound.J2}
\bal
%%%J_1 &\le C\eps \int_{|t| \le \cT_\eps} \eu^{-C|t|^{d/A}} \, dt \le C' \eps \,,
%%%\\
%%%%
J_2 &\le 2 (1 - \Phi(\sigma_\eps \cT_\eps)) \le \eps(- C \ln^2 \eps^{-1}) \le \eps^2 \,,
\eal
\ee
hence
$
\mu( I_\eps) \le  J_1 + J_2 \le C \eps \,,
$
under the condition \eqref{eq:cT.eps}. Considering $\eps^{-1} \ln^2\eps^{-1} = T$ as definition
of an implicit function $T \mapsto \eps$, and taking the logarithm of both sides, we see that
$\ln \eps(T) \asymp \ln T^{-1}$ as $T\to+\infty$, so \eqref{eq:cT.eps} would follow from
a more explicit condition
\be
\eps^{-1} \le T_N \, \ln^{-c} T \,, \quad c\in(0,+\infty) \,,
\ee
hence with $T_N \asymp N^A$, \eqref{eq:cT.eps} is fulfilled, whenever
\be
\eps \ge \heps(\theta,T_N) = N^{-A} \ln^{c} N = N^{-A\left( 1 - \frac{c \ln \ln N}{\ln N} \right)}
 \asymp N^{-\frac{A}{ 1 + \frac{c' \ln \ln N}{\ln N} } }  .
\ee
It suffices that, with an arbitrarily small $\theta>0$ and $N$ large enough, viz. $N \ge N_*(\theta)$,
\be
\label{eq:relation.eps.N}
\eps \ge N^{- \frac{A}{1+\theta}} ,
\ee
but actually here $\theta = \theta(N)=\ord{1}$ as $N\to\infty$.

\vskip1mm
\noindent
$\bullet$ \textbf{Bound on $J_1$.} Working with the sum $S_{M,N}(\om)$, we can make use only of
$\fa_n X_{n,k}(\om)$ with $M \le n \le N$. Fix $t\ne 0$, then for $T_M \le |t| \le T_N$, where
$L \mapsto T_L = C L^A$, we have, as before,
$$
\bal
\ln \big| \ffi_{S_{M,N}}(t)\big|^{-1} &\ge \sum_{n=N_t}^{N} K_n \ln \big| \ffi_{X}(\fa_n t)\big|^{-1}
\ge C t^2 \, \sum_{n=N_t}^{N} n^{d-1} \fa^2_n t
\\
&
\ge C_1 t^2 \, N_t^{-2A+d} \ge C_2 \, |t|^{d/A} \,.
\eal
$$
Thus
$$
\bal
J_1 =  \int_{ |t|\le T_M}  \big| \hchi_\eps(t) \, \ffi_S(t) \big| \, dt
+ \int_{ T_M\le |t|\le \cT_\eps}  \big| \hchi_\eps(t) \, \ffi_S(t) \big| \, dt
,=: J_1^- + J_1^+ \,.
\eal
$$
where for $J_1^-$ we can only use a trivial upper bound, replacing $|\ffi_{S_{M,N}}(t)|$ by $1$:
\be
\label{eq:bound.J1}
\bal
J_1^- &\le \int_{-T_M}^{T_M} \big| \hchi_\eps(t) \, \ffi_S(t) \big| \, dt
\le \int_{-T_M}^{T_M} \big| \hchi_\eps(t) \big| \, dt
\\
&
\le \eps \int_{-T_M}^{T_M}  \frac{ \big|\sin(4\eps t)\big|}{ \eps|t| } \eu^{ - \frac{\sigma_\eps^2 t^2}{2}} \big| \, dt
\le 2 \eps \int_{0}^{T_M} \frac{ \big| \sin(4\eps t) \big| }{ \eps t } \big| \, dt
\\
&
\le C_3 \eps \int_{1}^{T_M} \, dt < C_3\, T_M \eps \le C_4 M^{A} \, \eps
\,.
\eal
\ee

Collecting \eqref{eq:bound.J2} and \eqref{eq:bound.J1}
completes the proof of assertion (D) for the intervals $I = I_\eps$ of length $|I_\eps| \ge N^{- \frac{A}{1+\theta}}$:
$$
\mu_{S_{M,N}}(I_\eps) \le C_4 M^{A} \, \eps + \eps^2 \le C_5 M^{A}\, \eps \,.
$$
\qedhere

\vskip2mm

Clearly, the estimate (D) becomes efficient for $N \gg M$, in view of the restriction
\eqref{eq:relation.eps.N}.

\subsection{Auxiliary estimates for the characteristic functions}

The following simple inequality, easily proved by induction in $n\ge 1$, allows one to avoid
exponential moments in estimation of the characteristic functions of real-valued r.v. Once again,
it is to be stressed that the key bounds presented below are valid not only for a.s. bounded r.v.,
but in a substantially larger class of probability measures with just a few finite moments.
\ble
\label{lem:Taylor.exp} For any integer $n\ge 1$,
\be
\label{eq:lem.Taylor.exp}
\forall\, s\in\DR \qquad  \left| \eu^{\ii s} - \sum_{k=0}^{n} \frac{s^k}{k!} \right| \le \frac{|s|^{n+1}}{(n+1)!}.
\ee
\ele

\ble
\label{lem:bound.ln.ffi}
Assume that $\fm_3 := \esm{|X|^3} < \infty$ and let $\sigma^2 := \esm{X^2}$. Then
\be
 \forall\, |t| \le \frac{\sigma^2}{\fm_3} \qquad |1 - \ffi(t)| \le \half \sigma^2 t^2 < \half
\ee
and
\be
\label{eq:bound.l,.ffi}
\left| \ln \ffi(t) + \half \sigma^2 t^2 \right| \le \frac{1}{6} \fm_3|t|^3 + \frac{1}{4} \sigma^4 t^4
\le \frac{5}{12} \fm_3 |t|^3 \,,
\ee
and, consequently,
%%if $|t| \le \frac{3}{5} \sigma^{1/2}$, then
\be
\label{eq:bound.l,.ffi.2}
|t| \le \frac{3}{5} \sigma^{1/2} \quad \Longrightarrow \quad
\ln |\ffi(t)|^{-1} \ge \quart \sigma^2 t^2 \,.
\ee
\ele

\proof
From the moment inequality, valid whenever the moments involved are finite,
$$
\big(\esm{ |X|^a }\big)^{1/a} \le \Big(\esm{ |X|^b }\Big)^{1/b}\,, \;\; 0 < a \le b\,,
$$
it follows, by taking $a=2$ and $b=3$, that
$\sigma^6 \le \fm_3^2$.
Since $\esm{X}=0$, by \eqref{eq:lem.Taylor.exp} applied to
$\ffi(t) = \esm{ \eu^{\ii t X}}= 1 + \ii t \esm{X} + ...$ with $n=2$ and $s=tX$,
we have
$$
\forall\, |t| \le \frac{\sigma^2}{\fm_3} \qquad |1 - \ffi(t)| \le \half \sigma^2 t^2
\le \half \fm_3^{2/3} \left( \frac{1}{\fm_3^{1/3}}\right)^2 \,
 < \half.
$$
Next, by the Taylor expansion for the logarithm, valid whenever $|1 - \ffi(t)|<1$,
$$
\bal
-\ln \ffi(t) - \half \sigma^2 t^2
& =  1 - \ffi(t) - \half \sigma^2 t^2 + \sum_{k\ge 2} \frac{1}{k} \big( 1 - \ffi(t) \big)^k \,,
\eal
$$
and since $|1 - \ffi(t)| \le \half$, thus for all $k\ge 2$
$$
\frac{1}{k} \big| 1 - \ffi(t) \big|^k \le 2^{-k},
$$
the claim \eqref{eq:bound.l,.ffi} follows by summing the series $\sum_{k\ge 2} 2^{-k}$:
$$
\left| \ln \ffi(t) + \half \sigma^2 t^2 \right| \le \frac{1}{6} \fm_3|t|^3 + \frac{1}{4} \sigma^4 t^4
\le \frac{5}{12} \fm_3 |t|^3.
$$
Under an additional assumption $|t| \le \frac{3}{5} \sigma^{1/2}$, we obtain
$$
\bal
\ln |\ffi(t)|^{-1} &\ge \half \sigma^2 t^2 - \frac{5}{12} \fm_3 |t|^3
%%\\
%%%
%%&
\ge \half \sigma^2 t^2 \left( 1 - \frac{5\fm_3}{6 \sigma^2} |t| \right)
\\
&
\ge \half \sigma^2 t^2 \left( 1 - \frac{5\fm_3}{6 \sigma^2} \cdot \frac{3 \sigma^2}{5 \fm_3} \right)
= \quart \sigma^2 t^2 \,,
\eal
$$
owing to the same moment inequality as above, used again in the last line.
\qedhere

The above general results will be used in the situation where $X(\om) = a_{|x|} \om_x$
with $\pr{|\om_x|\le 1} = 1$ for all $x$, so
$\fm_3 = \esm{ a_{|x|}^3 |\om_x|^3} \le a_{|x|}^3 \, \mub_3$, where
\be
\label{def:mu.bar}
\mub_3 := \esm{|\om_\bullet|^3}\in(0,1] \,.
\ee
Notice that that the key ration used in Lemma \ref{lem:bound.ln.ffi} reads as
$$
\frac{\sigma^2}{\fm_3} = \frac{ \left( \esm{ a_{|x|}^2 |\om_x|^2} \right)^2 }{ \esm{ a_{|x|}^3 |\om_x|^3} }
= \frac{ \sigmab^2 }{ a_{|x|}^3\, \mub_3 }\, , \qquad
\sigmab^2 := \esm{|\om_\bullet|^2}\in(0,1] \,,
$$
and $\sigmab^2/\mub_3\in(0,+\infty)$ is a fixed parameter characterizing the common probability distribution of the
IID scatterer amplitudes $\om_x$. For example, $\sigma_b = \mub_3 = 1$ for the Bernoulli
distribution with values $\{-1, +1\}$.

\subsection{Thermal bath estimate for the cumulative potential}
\label{ssec:thermal.bath.ch.f}

\ble
\label{lem:thermal.bath.F.V}
Consider a random field $V(x,\om)$ on $\DZ^d$ of the form
$$
V(x,\om) = \sum_{y\in\DZ^d} \fu(y-x)\, \om_y \,,
$$
where $\fu$ is given by \eqref{eq:def.fu} and $\{\om_x, \, x\in\DZ^d\}$ are bounded IID r.v. with nonzero variance. Then the following holds true:

\begin{enumerate}[\rm(A)]
  \item The common characteristic function $\ffi_V(\cdot$ of the identically distributed r.v.
$V(x,\om)$, $x\in)\DZ^d$, obeys an upper bound
\label{lem:thermal.bath.bound}
\be
\all t\in\DR \quad \big| \ffi_V(t)\big| \le C \, \eu^{-|t|^{d/A}} \,.
\ee

  \item Consequently, the common probability distribution function $F_V(\cdot)$ of the
  cumulative potential at sites $x\in\DZ^d$ has the derivative $\rho_V \in\mcC(\DR)$.

  \item Let $v_* := \inf\, \supp \rho_V$, then $F_V(v_*+\lam) = \ord{|\lam|^\infty}$.
\end{enumerate}

\ele

\proof
The claim will be derived from the Main Lemma \ref{lem:Main},
so we only need to check the validity of its assumptions.
Denote
\be
\label{eq:def.cX.r.n.r}
\bal
\mcX_n &:= \{x\in\mcZ:\, |x|\in(\fr_{n-1}, \fr_n]\} \,, \quad n\in\DN,
\\
K_n & := \big| \mcX_n \big| \,,
\\
\fa_n &:= \fu(\fr_n) \equiv \fr_n^{-A}  \,.
\eal
\ee
The r.v. $X_{n,k}$ figuring in Lemma \ref{lem:Main} are now $\om_x$ with $x\in \mcX_n$, numerated
arbitrarily by $k\in[1, K_n]$.

Next, note that Lemma \ref{lem:bound.ln.ffi} applies here, since $\om_x$ are a.s. bounded, thus have finite absolute
moments of all orders.
Let
$N_t = \left\lceil C t^{1/A} \right\rceil$,
where $C$ is chosen so that for any $n \ge N_t$ one has
$$
\fa_{n} |t| \le \fa_{N_t} |t| \sim N_t^A \sim C^A t \le \frac{3}{5} \sigma^{1/2} \,,
$$
hence by Lemma
$$
 \ln \big| \ffi_\mu\big( \fa_n t \big) \big|^{-1} \ge  \frac{\sigmab^2 \,}{4} \fr_n^{-2A} t^2\,.
$$
Now the claim follows from Lemma \ref{lem:Main}:
$\big| \ffi_S(t) \big| \le \Const\, |t|^{-d/A}$.
\qedhere

\vskip2mm

For the proof of a fractional-exponential decay of the characteristic function at infinity, there
was no need to assess, in Main  lemma, the "ripple" sum $\mcS_1(t)$ which is in general a more delicate task.
However, for a particular (but rather rich) class of measures, a lower bound for $\mcS_1(t)$ that one can obtain
has the same order of magnitude
as the above one for $\mcS_2(t)$. Specifically, it suffices to assume the so-called Cram\`{e}r's
condition (a.k.a. Condition (C); cf. \cite{Cram70}) widely used in the
theory of asymptotic expansions for the limiting probability distribution (often probability density, in fact)
for the sums of IID r.v.

\ble
Let $\mu$ be a probability measure satisfying Cram\`{e}r's condition {\rm(C)}:
\be
\label{eq:cond.C}
\limsup_{|t|\to\infty} \, |\ffi_\mu(t)| \le \zeta < 1 .
\ee
Then the sum $\mcS_1(t)$ from  Lemma \ref{lem:Main} obeys for some $c>0$
$$
\mcS_1(t) \ge c \ln (\zeta^{-1}) \,  R_t = c_\zeta |t|^{d/A} \,.
$$
\ele

The proof is obvious, as each term in $\mcS_1(t)$ is trivially lower-bounded by $\ln (\zeta^{-1})$.

\vskip3mm
Without Cram\`{e}r's Condition (C), one needs in general more subtle equidistribution arguments
in order to show that, pictorially,
a typical term of the sum $\mcS_1$ brings a nonzero average contribution.
Quite fortunately, for the potentials $\fu(r) = r^{-A}$, this can be done with the help of
an elementary lemma due to P\'olya and Szeg\"{o}, as explained in the next Section \ref{ssec:Wintner}.

Observe that while Condition (C) is of course violated for
Bernoulli distributions, it is incomparably weaker than the Rajchman property (cf. \cite{Rajch22,Rajch28}),
i.e., the condition on the Fourier transform of a measure $\mu$,
$$
\lim_{|t|\to\infty} |\ffi_\mu(t)| = 0 \,,
$$
and the latter is substantially weaker
than the assumptions used by Campanino and Klein \cite{CamKle86}
and Klein \etal \cite{KleMarPer86}. In the latter work, the authors
expressed their hope that the conditions (1.1)--(1.2) from their Main Theorem (power-law decay at infinity
of the single-site characteristic function) would be sufficient for a regularity of the IDS
enabling one to prove Anderson localization in dimension higher than one with rather singular probability distribution
of the IID random potential. The question of whether this is true for the models on
higher-dimensional \emph{lattices} (or more general graphs) with short-range interaction potential
remains wide open and challenging, so it is rather curious to see how simple becomes
the analysis of regularity (and, as a result, of Anderson localization) in discrete models
under Cram\`{e}r's condition (C).

Although the onset of Anderson localization at low energies is no longer
an open problem for Anderson Hamiltonians in $\DR^d$, $d\ge 1$, with an alloy-type potential
and arbitrary nontrivial distribution of scatterers' amplitudes, owing to the deep works by
Bourgain and Kenig \cite{BK05}, Aizenman, Germinet, Klein and Warzel \cite{AGKW09}, and
Germinet and Klein \cite{GK13}, the proofs for short-range interaction potentials are rather complex,
as one can judge already by the considerable size of the paper \cite{GK13} summarizing the required techniques
and results achieved for arbitrarily singular single-site measures. Unless some new breakthrough is made
in this direction, it seems that any regularity weaker than log-H\"{o}lder continuity is essentially
as hard to treat as the Bernoulli case, indeed even harder, as evidence the efforts made in
\cite{AGKW09} precisely in order to extend the ideas and techniques by Bourgain and Kenig to singular
measures which are barely less singular than Bernoulli, yet  not exactly Bernoulli (possibly s.c.).

It seems, therefore, that in the framework of physically realistic, viz. nonlocal interaction potentials
the "regularity threshold", separating the "easy" models from those where the analysis of IDS and of
localization phenomena requires more involved methods, is brought much lower than for
the local interaction models: instead of the log-H\"{o}lder continuity, one can have a comfortable
setup under a significantly weaker Cram\`{e}r's condition.

\subsection{Comments on the proof of Assertion (A) of Main Lemma}
\label{ssec:Wintner}

The proof of the lower bound of the tidal sum $\mcS_2(t)$ was quite straightforward, but
the reader may wonder if the final estimate based on it is optimal, or it can be improved
by using $\mcS_1(t)$ instead. In general, assessing the "ripple" sum $\mcS_1(t)$ may  prove
to be a delicate task, requiring fine results on equidistribution properties of ergodic dynamical systems,
which in turn are related to algebraic properties of some parameters of those systems; see
a discussion in \cite{C16e}. However, in the particular case of the potentials admitting an asymptotic
$$
\fu(r) = r^\alpha F(r),
$$
where $F$ is a slowly varying function, viz. satisfies
$$
\forall\, c\in(0,+\infty) \qquad \lim_{r\to+\infty} \frac{F(cr)}{F(r)} = 1 \,,
$$
Wintner \cite{Wint1934} proved a lower bound
essentially equivalent (in notations of the present paper) to
$$
\mcS_1(t) \ge C F(t) \, |t|^{1/\alpha}.
$$
with the help of a result by P\'olya and Szeg\"{o}
\cite{PolSzego25}

\ble[P\'olya and Szeg\"{o}, \lcite{PolSzego25}{Section II.4.1, Problem 155}]
\label{lem:PolSze25}
Let be given a monotone increasing real sequence $\Br = (r_n)_{n\ge 1}$ such that its counting function
$$
N_t = N_t(\Br) := \sum_{n:\, r_n \le t} \equiv \card\myset{n:\, r_n \le t}
$$
is varying regularly with exponent $\lam>0$ at infinity, i.e., one has a representation
$N_t = t^\lam \csS(t)$ where $\csS(t)$ is a so-called slowly varying function at infinity:
$$
\forall\, c>0\quad \lim_{t\to\infty} \frac{\csS(ct)}{\csS(t)} = 1.
$$
Next, consider a
Riemann-integrable\footnote{This condition is important for the proof given in \cite{PolSzego25}.
The prototypical case is where $f$ is piecewise constant, and for the summation formula to be asymptotically sharp, the function
$f$ must admit two-sided bounds $\psi \le f \le \Psi$ by piecewise constant functions $\psi, \Psi$ with arbitrarily high
accuracy.}
function $f:\,(0,c]\to \DR$. Then
\be
\lim_{t\to+\infty} \frac{1}{N_t} \; \sum_{n:\, r_n\le t} f\left( \frac{r_n}{t} \right)
= C(\Br,f,\lam,c)
:= \int_0^{c^\lam}  f\left( s^{1/\lam} \right) \, ds\,.
\ee
\ele

Wintner used $f(s) = \ln \, \max\big[ \cos\left( s^{-1} \right), \, \half \big]$.
By identification of parameters used in  \cite{Wint1934} and here, one can see that
for power-law potentials $\fu$, the "ripple" sum
$\mcS_1(t)$ has the same order of magnitude as the tidal sum $\mcS_2(t)$.

Several problems formulated in \lcite{PolSzego25}{Section II.4.1} are close in spirit to the
equidistribution statements from the neighboring Sections II.4.2--II.4.5 in \cite{PolSzego25}.
It is therefore no surprise that Lemma \ref{lem:PolSze25} appears in Wintner's work
\cite{Wint1934} at the place where, for general potentials $\fu$, one would need
some sort of equidistribution arguments.

Notice that the fairly general Lemma \ref{lem:PolSze25} provides an accurate,
asymptotically sharp summation formula, but Wintner's application to the analysis
of characteristic functions proceeds by a one-sided bound, in order to fulfill the required
conditions. Specifically, taking $\max[\cdot\,, \, \half]$ in the definition of $f$,
one looses the benefit of small values of $|\cos(\cdot)|$, hence of large values of
$\ln \left(|\cos(\cdot)|^{-1}\right)$.

This ultimately leads to an upper bound for the characteristic function of
the infinite convolution measure at hand. In a subsequent paper on the same subject
\cite{Wint1935}, Wintner complemented his result by making use of the terms with
\emph{small} values of $|a_n t|$; the reader can see that his original argument from
the one-page paper \cite{Wint1935} is much more straightforward and does not require
any technical work and several approximations used in the proof of
\lcite{PolSzego25}{Section 2.4.1, Problem 155}. In Section \ref{ssec:thermal.bath.ch.f}
we merely adapt his elementary proof. Both estimates -- for $\mcS_1(t)$ and for
$\mcS_2(t)$ -- give qualitatively similar results, yet these are only upper bounds and not
an asymptotic formula.

\section{Infinite smoothness of the DoS and Wegner estimates}

\subsection{DoS in a thermal bath}
\label{ssec:inf.smooth.DoS.thermal.bath}

\proof[Proof of Theorem \ref{thm:DoS.xi.Lam}]
The claim follows easily from the Main Lemma \ref{lem:Main}; we only need to identify
the key ingredients  of the latter:
$$
\bal
\mcX_n &: = \myset{x\in\DZ^d:\, \dist\left(x, \Lam_L\right) \in[\fr_n, \fr_{n+1}) }\,,
\;\; K_n := \big|\mcX_n \big| \,,
\\
\myset{ \om_x, \, x\in\mcX_n } & \leftrightarrow \myset{ X_{n,k}, \, k=1, \ldots, K_n }
\\
M &:= L, \;\; N = +\infty\,,
\\
S_{M,N}(\om) &= \sum_{n=M}^\infty \sum_{k=1}^{K_n} \fa_n X_{n,k} \equiv \sum_{x:\, |x|\ge L} \fu(|x|) \om_{x}
%%\\
\eal
$$

In some situations, one may want for technical reasons to take instead of $\mcX_n$ only a suitable
subset thereof, and include the potential induced by the remaining amplitudes $\om_x$ in
a r.v. $Y(\om)$. Then the latter can be effectively ignored in the calculations, without invalidating them,
by conditioning first on $Y$, i.e. on the unused random potentials. All one needs is that the cardinality
of the reduced $\mcX_n$ remain sufficiently large for the desired final bound. To be more precise,
the conclusion on infinite smoothness remains valid as long as $|\mcX_n| \ge C n^{\kappa}$ for
some $\kappa>0$, no matter how small; it would only affect the decay exponent $a(\kappa,A)>0$ in
$|\ffi(t)|\le \eu^{ - C |t|^{a(\kappa,A)}}$.

By assertion (B) of Lemma \ref{lem:Main},
$$
\bal
\all t \in\DR \quad
\big| \ffi_{\xi_\Lam}(t)\big| \le \eu^{- C \, M^{-2A+d}|t|^{d/A}} \,.
\eal
$$
This implies the existence of the probability density $\rho_{\xi_\Lam}\in\mcC^\infty(\DR)$
of the measure $dF_{\xi_\Lam}(E)$. Due to the representation \eqref{eq:H.H0.xi}
with scalar random operator $\xi_\Lam(\om) \one_\Lam$, one can label all EVs $\lam_j(\om)$ of
$H_\Lam(\om)$ in a measurable way so that
$$
\lam_j(\om) = \hlam_j(\om) + \xi_\Lam(\om), \;\; j=1, \ldots, |\Lam|,
$$
where all r.v. $\hlam_j$ are independent of $\xi_\Lam$, so the infinite smoothness
of the IDS (hence, of the DoS) in $\Lam$ easily follows.
\qedhere

\subsection{Wegner estimates}
\label{ssec:Wegner}

Aiming to the applications to Anderson localization, we now have to operate
with a restricted, annular "bath" of finite size, the complement of which is "frozen".
This is necessary for obtaining a satisfactory replacement for the IAD property
very valuable in the short-range interaction models.

In the following theorem appear two important parameters, $\theta>0$ which has the same meaning as in
Main Lemma \ref{lem:Main} (cf. \eqref{eq:relation.eps.N}), and $\tau>1$, which will be used in Section
\ref{sec:loc.polynom} and can be chosen arbitrarily large.

\proof[Proof of Theorem \ref{thm:Wegner}]
The required bound follows from assertion (C) of Lemma \ref{lem:Main} which was tailored specifically
to suit the Wegner bound in a finite annular "bath". By translation invariance of the random field
$\bm{\om} = (\om_x)_{x\in\DZ^d}$, we can assume w.l.o.g. that $u=0$.

Identification of the principal ingredients of
Lemma \ref{lem:Main} is as follows:
$$
\bal
\mcX_n &: = \myset{x\in\DZ^d:\, |x| = n }\,, \;\; K_n := \big|\mcX_n \big| \,,
\\
\myset{ \om_x, \, x } & \leftrightarrow \myset{ X_{n,k}, \, k=1, \ldots, K_n }
\\
M &:= L, \;\; N = R_L = R_L(\tau)\,,
\\
S_{M,N}(\om) &= \sum_{n=M}^N \sum_{k=1}^{K_n} \fa_n X_{n,k} \equiv \sum_{x:\, |x|\in[L, R_L]} \fu(|x|) \om_{x}
\\
Y(\om) &= \sum_{n=0}^N \sum_{k=1}^{K_n}  \fa_n X_{n,k} \equiv \sum_{x:\, |x| < L } \fu(|x|) \om_{x} \,.
\eal
$$
Proceeding as in Theorem \ref{thm:DoS.xi.Lam}, we obtain the representation
\be
\label{eq:thm.Wegner.frozen.xi}
H_\ball(\om) = \widetilde{H}_\ball(\om) + \xi_\ball(\om) \, \one_\ball\,,
\ee
where the random operator $\widetilde{H}_\ball(\om)$ is independent of the r.v. $\xi_\ball(\om)$,
and the latter is generated from the amplitudes $\om_x$ with $x$ in the annulus
$\cA = \ball_{R_L}(0) \setminus \ball_{L}(0)$, modulating the "plateaus" of the respective
interaction potentials $y\mapsto \fu(|y-x|)$, $y\in\ball_L(0)$, covering entirely the ball
$\ball = \ball_L(0)$.
By Lemma \ref{lem:Main}, $\xi_\ball$ fulfills, for any interval $I$ of length
\be
\label{eq:rel.eps.L}
\eps_L = N^{-\frac{A}{1+\theta}} \equiv L^{ -\frac{A\tau}{1+\theta}}
\ee
the concentration estimate (cf. \eqref{eq:Main.Lemma.D})
\be
\pr{ \xi_\ball(\om) \in I_\eps } \le C M^A\, |I_\eps| \equiv C\, L^A \, \eps
\,,
\ee
with $\theta>0$ arbitrarily small, provided $M = L$ is large enough.

Now we make use of this freedom and pick $\theta\in(0, \tau - 1)$, then solving \eqref{eq:rel.eps.L}
for $L$ as implicit function of $\eps_L$, we see that
\be
\pr{ \xi_\ball(\om) \in I_\eps } \le C M^A\, |I| \equiv C\, L^A \,
\eps_L^{ 1 - \frac{1+\theta}{\tau}}
= \eps_L^{ 1 - \frac{1+\theta}{\tau}} = \eps_L^{ \beta}
\,,
\ee
where
\be
0 < \beta = 1 - \frac{1+\theta}{\tau} \;  \tto{\tau\to+\infty} \, 1\,.
\ee
This proves the EVC estimate \eqref{eq:cor.Wegner.bound}, since $H_\ball(\om)$
acts in the Hilbert space $\ell^2(\ball)$ of dimension $|\ball|$.
\qedhere

\bre
\label{rem:Wegner.discrete.continuos}
It is readily seen that the same argument proves a direct analog of
the EVC estimate \eqref{eq:cor.Wegner.bound} for the Anderson Hamiltonians in $\rL^2(\ball)$,
$\ball\subset\DR^d$,
with various boundary conditions, due to the crucial decomposition
\eqref{eq:thm.Wegner.frozen.xi} which has exactly the same form in discrete and continuous
configuration spaces; idem for the Anderson Hamiltonians on quantum graphs. The only difference
comes from the Weyl asymptotics for the EVs, responsible for an extra factor in the RHS depending
upon the position of $I$ (say, of length $|I|\le 1$) in the energy axis.
\ere

\section{ILS estimates at low energies via "thin tails"}
\label{sec:ILSthin.tail}

Due to non-negativity of the Laplacian $H_0$,
%\be
\begin{align}
\label{eq:thin.tail.1}
\pr{ E_0^\Lam(\om) < \lam } & \le \pr{ \min_{x\in\Lam} V((x,\om) < \lam }
\\
\label{eq:thin.tail.2}
&
\le |\Lam| \, \min_{x\in\Lam} \; \pr{ V((x,\om) < \lam }
\\
\label{eq:thin.tail.3}
&
\le |\Lam| \; F_{V(x,\cdot)}(\lam) = |\Lam| \; \ord{ \lam^\infty } \,,
\end{align}
%\ee
where the last equality is due to assertion (C) of Lemma \ref{lem:thermal.bath.F.V}.
This strong form of decay of the EV distribution
at the bottom of spectrum is, however, unsuitable for the application to the ILS (initial length scale)
estimate in the course of the MSA, for it lacks independence in the setting where several volumes
$\ball_{L_0}(u_j)$, $j=1, 2, \ldots, n$ are considered simultaneously. This can be remedied as follows.

Given a ball $\ball = \ball_{L_0}(u)$, denote $\ball^+ = \ball_{2L_0}(u)$, and for any given sample
$\om\in[0,1]^{\DZ^d}$, introduce a sample measurable with respect to a sub-sigma-algebra
$\fF_{\ball^+} = \fS[\om_y, \, y\in\ball^+]$:
\be
\om^+_x =
\left\{
  \begin{array}{ll}
   \om_x , & \text{ if $x\in\ball^+$,} \\
    1 \equiv \sup \, \supp \mu, & \text{otherwise .}
  \end{array}
\right.
\ee
Then, obviously,
$$
\forall\, x\in\ball \quad V(x,\om) < V(x,\om^+).
$$
Depending on the reader's point of view, the above inequality can be understood as holding a.s.
(a traditional probabilists' superstition: never to be "sure" but at best "almost sure"),
or everywhere -- if the basic probability space is defined from the very beginning as
$[0,1]^{\DZ^d}$ rather than $\DR^{\DZ^d}$ (analysts are likely to prefer this variant).
Just like $\om^+$, the modified potential $V(\cdot,\om^+)$ is $\fF_{\ball^+}$-measurable.
Therefore, for any family of balls $\ball_{L_0}(u_i)$, $i=1, 2, \ldots$, with
pairwise distances $|u_i - u_j| > 4 L_0$, and any
measurable functionals of $V\big|_{\ball^+(u_i)}$, the family of random variables
$$
\zeta_i(\om) = f_i\left(V\left(\cdot, \om^{(+, i)}\right) \big|_{\ball^+(u_i)} \right)
$$
is independent. Here $\om^{(+, i)}$ is obtained from $\om$ in the same way as $\om^+$
for a given ball $\ball_{L_0}(u)$ in the general construction explained above, hence
the samples $\myset{ \om^{(+,i)}, \, i=1, 2, \ldots}$ are independent, yielding the same
property for the r.v. $\zeta_i$.

\btm
Fix any $L_0>1$ and consider the Hamiltonian $H_{\ball_{L_0}(u)}(\om)$ with an arbitrary $u\in\DZ^d$.
Assume that the interaction potential decays as $\fu(r) = r^{-A}$, $A>d$, and introduce a larger ball
$\ball^+=\ball_{\fc L_0}(u)$ and the sigma-algebras
\begin{itemize}
  \item $\fF_{\ball^+}$ generated by all scatterers' amplitudes $\om_y$ with $y\in \ball^+$,

  \item $\fF_{\ball^+}^\perp$ generated by all scatterers' amplitudes $\om_y$
  with $y\in\DZ^d\setminus \ball^+$.
\end{itemize}
Then for any $\theta\in(0,1)$ there exists some $C_\theta>0$ such that
\be
\pr{  E_0^\Lam(\om) \le L_0^{-\theta} } \le  \eu^{ - C_\theta L_0^d } \,.
\ee

\etm

\proof
%%We will derive the claim from a stronger bound off the form \eqref{eq:thin.tail.2} following
%%from a modified variant of \eqref{eq:thin.tail.3}.
%%
Fix any $x\in\ball_{L_0}(u)$, let $R_0 = L_0^{\theta}$, $\theta>0$, and denote
$Q_x = \ball_{2R_0}(x)\setminus \ball_{R_0}(x)$; note that $|Q_x| > R_0^d$. For any
$\kappa>0$,
$$
\max_{y\in Q_x} \, \om_y > \kappa \;\; \Rightarrow \;\; V(x,\om) \ge C_A \kappa R_0^{-A}
> L_0^{-\theta} \,,
$$
whence
$$
\bal
\forall\, x\in\ball_{L_0}(u) \qquad  \pr{ V((x,\om) < \lam }
& \le \pr{ \forall\, y\in Q_x \;\; \om_y \le \kappa }
\\
&
= \prod_{y\in Q_x} \pr{\om_y \le \kappa } = \eps_\kappa^{ |Q_x| }
\\
&
= \eu^{ - \ln\eps_\kappa^{-1} R_0^d }
\le \eu^{ - C(\eps_\kappa,d) L_0^d }
\eal
$$
yielding
$$
  \pr{\min_{x\in\ball_{L_0}(u)} \, V((x,\om) < \lam }
  \le \big|\ball_{L_0}(u)\big| \; \eu^{ - C(\eps_\kappa,d) L_0^d }
  \le \eu^{ - C' L_0^d }\,,
$$
for some $C' = C'(\kappa, d)>0$. By non-negativity of the kinetic energy operator $H_0$,
this proves the claim (cf. \eqref{eq:thin.tail.1}.
\qedhere

The first scenario leading to the onset of Anderson localization is more universal and robust than the one considered
in the next subsection; here we do not make any assumption on the magnitude of the potential and do not
attempt to achieve a global bound on the entire spectrum (which is usually possible in discrete systems
and/or in one dimension). This will result in ILS estimates easily adapted to the continuous alloy models
in $\DR^d$, $d\ge 1$, as well as in a large class of quantum graphs, with tempered underlying
combinatorial graphs of coupling vertices.

%%%\ble
%%%
%%%\ele
%%%
%%%\proof
%%%
%%%\qedhere

\section{ILS estimates under large disorder}
\label{sec:ILS.strong.disorder}

\subsection{Individual EV concentration estimates}

We start with the weakest estimate which has, on the other hand, the general form
closest to the usual strong-disorder ILS bound used in MSA. It suits to the scaling analysis
based on scale-free probability estimates at the initial scale (cf. \cite{Sp88,GK01}).

An ILS bound is most efficiently used via the Combes--Thomas estimate \cite{CT73}, and the reader
familiar with this technique can see that in applications to the MSA,
it suffices to set $\eps=1$ (or any other fixed positive number)
in the inequality \eqref{eq:thm.ILS.one.EV}.

\btm
Let $\Sigma_{\ball, g}(\om) = \Sigma\big(H_\ball(\om)\big) = \myset{ E_j^{(\ball,g)}(\om), \, j=1, \ldots, |\ball|}$
be the set of random eigenvalues of $H_{\ball,g}(\om) = -\Delta_{\ball} + gV(\om)$, numbered
in a measurable way, counting multiplicity. Then
for any $\eta\in(0,1]$ and $\eps>0$ there  exist $L_0\in\DN$ and $g>0$ such that
\label{thm:ILS.one.EV}
\be
\label{eq:thm.ILS.one.EV}
 \sup_{E\in\DR} \;
\Bigpr{ \dist\Big(\Sigma_{\ball,g}(\om), E \Big) \le \eps  } \le \eta.
\ee
\etm

\proof
We will actually establish a slightly stronger bound, with the scale-free threshold
$\eta>0$ replaced with
$L_0^{-\kappa}$ with some $\kappa\in(0,A-d)$.

By Main Lemma \ref{lem:Main}, for any $x\in\ball$ and any interval $I_\delta$ of length
$\delta \ge L^{-A+\theta}$ with arbitrarily small $\theta>0$ and $L$ large enough (depending on $\theta$),
$$
\bal
\pr{ V(x,\om) \in I_\delta } \le C \, \delta,
\eal
$$
hence with $\delta = L^{-A+\theta}$, $\theta := A-d-\kappa$, $\kappa\in(0, A-d)$,
and sufficiently large $L$,
\be
\label{eq:prob.I.delta}
\bal
\pr{ \exists\,x\in\ball: \;  V(x,\om) \in I_\delta } \le C \,L^d \delta
= C \,L^{-A+ \theta +d} < L_0^{-\kappa}
\,.
\eal
\ee

Fix $\eps>0$ and let $\om$ be such that $\dist\big(\Sigma_{\ball, g}(\om), E\big) < \eps$, then
there exists at least one EV $E_i^{(\ball,g)}(\om)\in J_\eps(E) = (E-\eps, E+\eps)$. Since
$H_{\ball,g}(\om) = gV(\om) -\Delta_{\ball}$ and $\| \Delta_{\ball} \| \le 4d$,
where the EVs of  $gV(\om)$ form the set $\myset{gV(x,\om), \, x\in\ball}$, there exists at least
one $x\in\ball$ such that
$$
|gV(x,\om) - E| \le \eps + 4d \; \Longrightarrow \;
|V(x,\om) - g^{-1}E| \le \frac{\eps + 4d}{|g|}
$$
Equivalently, $V(x,\om) \in I_\delta := [g^{-1}E - \frac{\delta}{2}, \, g^{-1}E + \frac{\delta}{2}]$
with
$$
\delta = \delta(g) = \frac{\eps + 4d}{|g|}.
$$
Let $|g| = (\eps + 4d)L^{-A+\theta}$, then $\om$ must be contained in the event
$$
\mcR_\delta := \myset{ \exists\,x\in\ball: \;  V(x,\om) \in I_\delta },
$$
with $\pr{\mcR_\delta} < L_0^{-\kappa}$, according to \eqref{eq:prob.I.delta},
so the claim follows by taking $L_0 \ge \eta^{-1/\kappa}$.
\qedhere

\vskip3mm

The next result gives a stronger, power-law decay of the ILS probability estimate for the
MSA induction. The price to pay for this improvement is a possibly large "isolation zone"
around the balls $\ball_{L_0}(u_j)$, $j=1, \ldots, S$, in the scaling scheme where
$S\ge 1$ bad balls should be tolerated in the decay analysis of the Green functions
(cf. \cite{GK01}).

\btm
Let $\Sigma_{\ball, g}(\om) = \Sigma\big(H_\ball(\om)\big) = \myset{ E_j^{(\ball,g)}(\om), \, j=1, \ldots, |\ball|}$
be the set of random eigenvalues of $H_{\ball,g}(\om) = -\Delta_{\ball} + gV(\om)$,
numbered in a measurable way counting multiplicity.
Given $R\in\DN$, denote by $\fF_R$ the sigma-algebra generated by the amplitudes
$\om_x$ with $|x|>R$, given $\fF_R$.
Fix any $b>0$.
%%  and let $R = L^\tau$ with $\tau>\frac{d+b}{A}$.
Then for any (viz., arbitrarily large) $\eps>0$ there  exist $L_0\in\DN$ and $g>0$ such that
\label{thm:ILS.one.EV}
\be
\label{eq:thm.ILS.one.EV}
 \sup_{E\in\DR} \;
\pr{ \dist\Big(\Sigma_{\ball,g}(\om), E \Big) \le \eps \cond \fF_R }
%%\le L_0^d R^{-A+\theta}
\le L_0^{-b}.
\ee
\etm

\proof
Denote for
brevity\footnote{In other words, computing the probability $\prsub{R}{\cdot}$,
we can make a good use of the random fluctuations of the cumulative potential $V$ coming from $\om_x$
with $x\in\ball_R(u)$, while the remaining amplitudes generate a background potential considered as
"frozen", thus useless for the regularity analysis.}
$\prsub{R}{\cdot} \equiv \pr{\cdot \, \cond \fF_R}$.
By assertion (D) of Main Lemma \ref{lem:Main}, with $M=1$, for any $x\in\ball$ and any interval $I_\delta$
with $|I_\delta| =\delta \ge R^{-A+\theta}$, arbitrarily small $\theta>0$ and $L$ large enough (depending on $\theta$),
$$
\bal
\prsub{R}{ V(x,\om) \in I_\delta } \le C \, \delta,
\eal
$$
hence with
%%$$
\begin{align}
\label{eq:thm.Wegner.2.R}
R &= L^{\tau}, \;\; \tau>\frac{b+d}{A} \,,
\\
\label{eq:thm.Wegner.2.delta}
\delta &= R^{-A+\theta} \,,
\\
\label{eq:thm.Wegner.2.theta}
\theta &:= A-d-\kappa, \;\; \kappa\in(0, A-d)\,,
\end{align}
%%$$
and sufficiently large $L$,
\be
\label{eq:prob.2.I.delta}
\bal
\prsub{R}{ \exists\,x\in\ball: \;  V(x,\om) \in I_\delta } \le C \,L^d \delta
= C \,L^{-\tau A+ \theta +d} < L_0^{-b }
\,.
\eal
\ee

Fix $\eps>0$ and let $\om$ be such that $\dist\big(\Sigma_{\ball, g}(\om), E\big) < \eps$, then
there exists at least one EV $E_i^{(\ball,g)}(\om)\in J_\eps(E) = (E-\eps, E+\eps)$. Since
$H_{\ball,g}(\om) = gV(\om) -\Delta_{\ball}$ and $\| \Delta_{\ball} \| \le 4d$,
where the EVs of  $gV(\om)$ form the set $\myset{gV(x,\om), \, x\in\ball}$, there exists at least
one $x\in\ball$ such that
$$
|gV(x,\om) - E| \le \eps + 4d \; \Longrightarrow \;
|V(x,\om) - g^{-1}E| \le \frac{\eps + 4d}{|g|}
$$
Equivalently, $V(x,\om) \in I_\delta := [g^{-1}E - \frac{\delta}{2}, \, g^{-1}E + \frac{\delta}{2}]$
with
$$
\delta = \delta(g) = \frac{\eps + 4d}{|g|}.
$$
To satisfy \eqref{eq:thm.Wegner.2.delta},
let $|g| = (\eps + 4d)L^{-A+\theta}$, then $\om$ must be contained in the event
$$
\mcR_\delta := \myset{ \exists\,x\in\ball: \;  V(x,\om) \in I_\delta },
$$
with $\pr{\mcR_\delta} < L_0^{-b}$, according to \eqref{eq:prob.2.I.delta}.
This proves the claim.
\qedhere

\section{Proof of localization}
\label{sec:loc.polynom}

For brevity, we concentrate on the case of strong disorder and assume that the GFs
on some initial scale $L_0$ decay exponentially, with exponent $m_0 = \big(1+L_0^{-1/8}\big)m$,
$m\ge 1$. It is well-known that only a minor adaptation is required in the case of
"extreme energies", near spectral edge(s), where the the decay exponent $m_0$ may be as small as
$m_0 = L_0^{-\theta}$, $\theta\in(0,1)$; cf., e.g., \lcite{GK01}{Section 5.4}),
\lcite{St01}{Section 3.2}).

It was shown in Section \ref{sec:ILSthin.tail} that such a bound can be established with the help
of the "thin tails" argument, replacing its "Lifshitz tails" counterpart, with high
probability, viz. $p_0 \ge 1 - \eu^{-c_1 L_0^c}$, $c, c_1>0$, which is more than sufficient to
start the MSA induction. The path laid down in the spectral theory of random
operators several decades ago was based on the observation that IID or
IAD (Independent At Distance) random
potentials in Anderson-type Hamiltonians lead to a simplification of \emph{mathematical} analysis
of the localization problem; obviously, such a simplification of the model was in contradiction
with the physical reality where interactions have infinite range. It was only later, in 2005,
that one fully assessed the level of mathematical difficulties brought up by that simplification
in a general context of singular marginal distributions of the underlying disorder in discrete
models used in physics for modeling continuous solid-state systems in the so-called tight-binding
approximation which, curiously, was intended for making analysis \emph{simpler}.
In essence, this paper addresses a \emph{different} problem of
mathematical physics, which is closer to the physics and, quite fortunately, in many
aspects less hard on the technical level.

\vskip1mm
\noindent
$\blacklozenge$ Needless to say, the crucial fact that makes the usual MSA machinery to work here is Theorem
\ref{thm:Wegner} establishing a comfortable Wegner estimate. This is precisely what allows one
to avoid a radical re-writing the MSA induction, performed by Bourgain and Kenig \cite{BK05}
for short-range potentials with Bernoulli distribution of the local random amplitudes, and
and by Germinet and Klein \cite{GK13} for arbitrary nontrivial IID random amplitudes (again, for
short-range potentials).

\subsection{Deterministic analysis}

We adapt the strategy from \cite{KirStoStolz98a}.

Working with a Hamiltonian $H_{\ball_L(u)} = -\Delta_{\ball_L(u)} + gV$
in a given ball $\ball_L(u)$, it will be necessary to know the values of the amplitudes $\om_y$
with $y$ in a larger ball $\ball_{R_L}(u)\supset\ball_L(u)$, where the specific choice of $R_L$
depends upon the decay rate $r \mapsto r^{-A}$ of the interaction potential $\fu(r)$, along with
some other parameters of the model and of the desired rate of decay of EFCs to be proved.
Below we set $R_L = L^\tau$, $\tau>1$.

\bde
\label{def:NS.NR.polynom}
Let be given a ball $\ball=\ball_L(u)$.
A configuration $\Bfq\in \BfQ_{\DZ^d}$
is called
\begin{enumerate}[\rm(1)]
  \item $(E,\eps,\ball)$-non-singular iff the resolvent $G_{\ball}(E)$ of the operator
$$
H_{\ball_L(u)} = -\Delta_{\ball_L(u)} + \BU[\Bfq]\big|_{\ball_{L}}
$$
  (cf. the definition of $\BU[\Bfq]$ in \eqref{eq:def.BU.2}) is well-defined and satisfies
\be
\label{eq:def.NS}
\max_{x\in \ball_{L/3}(u)} \; \max_{y\in \pt^-\ball_{L}(u)}
 \; \left\| G_{\ball}(x,y; E) \right\| \le \eps\,;
\ee

  \item $(E,\gamma,\ball)$-non-resonant iff
\be
\label{eq:def.NR}
  \dist\big( \Sigma( H_{\ball} ), \, E \big) \ge \gamma.
\ee
\end{enumerate}
\ede

When the condition \eqref{eq:def.NS} (resp., \eqref{eq:def.NR}) is violated,
$\Bfq$ will be called $(E,\eps)$-singular (resp., $(E,\gamma)$-resonant).
We will be using obvious shortcuts $(E,\eps,\ball)$-NS, $(E,\eps,\ball)$-S, $(E,\gamma,\ball)$-NR and
$(E,\gamma,\ball)$-R.

\bde
\label{def:SNS.SNR.polynom}
Let be given a ball $\ball=\ball_L(x)$ and a real number $\tau>1$.
A configuration $\Bfq_{\ball_{L^\tau}}\in\BfQ_{\ball_{L^\tau}}$
is called
\begin{enumerate}
  \item $(E,\eps,\ball)$-SNS (strongly non-singular, or stable non-singular) iff
   for any configuration of amplitudes $\Bfq_{\ball^\rc}\in \BfQ_{\ball^\rc}$
   the extension of $\Bfq_{\ball_{L^\tau}}$ to the entire lattice,
   $\Bfq = (\Bfq_{\ball_{L^\tau}},\Bfq_{\ball_{L^\tau}^\rc})$ is $(E,\eps,\ball)$-NS;

  \item $(E,\gamma,\ball)$-SNR (strongly NR, or stable NR)
   iff for the configuration  $\BfQ_{\ball^\rc} \ni \Bfq_{\ball^\rc} \equiv 0$
   the function
   $V_\ball = \BU[\Bfq_{\ball_{R_L}} + \Bfq_{\ball_{R_L}^\rc}]\big|_{\ball_{R_L}}
    = \BU[\Bfq_{\ball_{R_L}}]\big|_{\ball_{R_L}}$ is
   $(E, \gamma)$-CNR.
\end{enumerate}
\ede

In subsection \ref{ssec:scaling} we work in the situation where the potential $V:\DZ^d\to\DR$ is fixed,a
and perform a deterministic analysis of finite-volume Hamiltonians.
It will be convenient to use a slightly
abusive but fairly traditional terminology and attribute the non-singularity and non-resonance
properties
to various balls $\ball$ rather than to a configuration $\Bfq$ or a cumulative potential $V =
\BU[\Bfq]$,
which will be fixed anyway. Therefore, we will refer, for example, to $(E,\eps)$-NS balls
instead of $(E,\eps,\ball)$-NS  configurations $\Bfq$. Similarly, we use the notions
of $(E,\gamma)$-NR, $(E,\eps)$-SNS or $(E,\gamma)$-SNR \emph{balls}.

\subsection{Scaling scheme}
\label{ssec:scaling}
Fix $\DN\ni d\ge 1$, $A>d$ and the interaction potential $\fu(r) \;\big(\sim r^{-A} \, \big)$ of the form
\eqref{eq:def.fu}.
Further, fix an arbitrary number $b>d$, which will represents the desired polynomial decay rate of the key
probabilities in the MSA induction, and let
\begin{align}
\label{eq:con.alpha.tau.S}
\alpha & >  \tau > \frac{b}{A-d } \,, \quad
\DN \ni S > \frac{b\alpha}{b - \alpha d} \,, \quad
L_{k+1} = \big\lfloor L_k^{\alpha} \big\rfloor \,,  \;\; k\ge 0 \,,
\end{align}
with $L_0$ large enough, to be specified on the as-needed basis. A direct analog of the well-known
deterministic statement \lcite{DK89}{Lemma 4.2} is the following statement adapted to long-range interactions
essentially as in \cite{KirStoStolz98a}. Introduce useful notation:

\be
\label{eq:def.eps.k.AL}
%%\eps_k := \eu^{ - m_k L_k }\,, \;\;
m_k := \left(1 + L_k^{-1/8}\right)m \,, \;\;
\eps_k := 4 L_k^{-\tau A + \theta} \,, \;\;
\theta\in(0,1)\,.
\ee
The value of $\theta$ can be chosen as small as one pleases.

\ble[Conditions for strong non-singularity]
\label{lem:cond.SNS.polynom}
Consider a ball $\ball=\ball_{L_{k+1}}(u)$, $k\ge 0$, and suppose that
\vskip1mm
\par\noindent
{\textnormal{\textbf{(i)}}} $\ball$ is $(E, \eps_k)$-SNR;
\vskip1mm
\par\noindent
{\textnormal{\textbf{(ii)}}} $\ball$ contains no collection of balls $\{\ball_{L_k}(x_i), 1\le i \le S+1\}$,
with pairwise $2L_k^{\tau}$-distant centers, neither of which is $(E, \eu^{-m_k L_k})$-SNS.
\par\noindent
Then $\ball$ is $(E,m)$-SNS.
\ele
\proof
Derivation of the NS property can be done essentially in the same way as in \cite{DK89} and in numerous
subsequent papers, with minor adaptations. See for example
\lcite{C16b}{proof of Lemma 7} where the singular balls
are also supposed to be pairwise $L_k^\tau$-distant, $\tau>1$. This does not actually require any significant
modification of the original argument from the work by von Dreifus and Klein \cite{DK89}. What is important here, is
that $(E,m)$-NS property of $\ball_{L_{k+1}}(u)$ (not the \emph{strong} NS) is derived from
weaker versions of the hypotheses \textbf{(i)}--\textbf{(ii)}, where SNS and SNR conditions are replaced
with their NS and NR counterparts. The proof refers only to the potential $V$ in $\ball_{L_{k+1}}(u)$,
which is fixed, so the notion of stability (hence variation of $V$) just does not appear in the proof.

To show that the  \emph{strong} (stable) non-singularity property also holds true, one can use induction on scales $L_k$.
We have to show that the NS property of the larger ball $\ball_{L_{k+1}}(u)$ is stable with respect to
arbitrary fluctuations of the random amplitudes $\om_y$ with $y\not\in \ball_{L^\tau_{k+1}}(u)$.
According to what has just been said in the previous paragraph, it suffices to check the stability of the
properties
\vskip1mm
\noindent
$\mathbf{(i')}$
$\ball_{L_{k+1}}(u)$ is $(E, \gamma_k)$-NR,

\vskip1mm
\noindent
$\mathbf{(ii'')}$ \, $\ball_{L_{k+1}}(u)$ contains no collection of balls $\{\ball_{L_k}(x_i), 1\le i \le S+1\}$,
with pairwise $2L_k^{\tau}$-distant centers, neither of which is $(E,m)$-NS,

\vskip1mm
\noindent
under the hypotheses \textbf{(i)}--\textbf{(ii)}.

There is nothing to prove for the stability of $\mathbf{(i')}$, as it is asserted by $\mathbf{(i)}$.

Stability of the NS property of balls $\{\ball_{L_k}(x_i)$ can be derived recursively, with the help of
the arguments from \cite{DK89}, \lcite{C16b}{proof of Lemma 7}. The non-singularity property
(even its conventional, non-stable variant) is quite implicit; in fact, the whole
point of making it a sort of "black box" in \cite{DK89} was a realization that it is extremely
difficult, if realistic at all, to trace the unwanted events, susceptible to prevent the onset of
localization in a ball at some induction step $k\gg 1$, down to the multitude of smaller balls
of size $L_0$ inside $\ball_{L_{k}}(x)$. Since it is derived inductively, we have to make sure that
the fluctuations of $\om_y$ with $y\not\in \ball_{L^\tau_{k+1}}(u)$ cannot destroy the non-resonance and
non-singularity properties in the relevant balls $\ball_{L_j}(x)\subset\ball_{L_{k+1}}(u)$,
$j=0, \ldots, k$.

On the scale
$L_0$ the non-singularity is derived from non-resonance, with a comfortable gap between an energy $E$
and the spectrum in the ball of radius $L_0$, which provides the base of induction.
Evidently, given any ball
$\ball_{L_{j}}(x) \subset \ball_{L_{k+1}}(u)$ one has
$$
\all j=0, \ldots, k \quad \ball^{\rc}_{L_{k+1}}(u) \subset \ball^{\rc}_{L_{j}}(x) \,.
$$
In other words, stability encoded in the SNS or SNR properties of smaller balls
$\ball_{L_{j}}(x) \subset \ball_{L_{k+1}}(u)$ is stronger than what is required for the stability
w.r.t. fluctuations $\om_y$ outside a much larger ball $\ball_{L^\tau_{k+1}}(u)$.
We conclude that the claim follows indeed from the the hypotheses \textbf{(i)}--\textbf{(ii)}.
\qedhere

\subsection{Probabilistic analysis}

It follows directly from Definition \ref{def:SNS.SNR.polynom} that any event of the form
$$
 \mcA\big(\ball_L(x), E, m \big) = \myset{ V_\fq(\cdot\,; \om)\big|_{\ball_L(x)} \text{ is $(E,m)$-SNS }  }
$$
is measurable w.r.t. the sigma-algebra $\fF^\fq_{\ball^\tau_{L}(x)}$.
%
%%Indeed, the potential induced
%%on $\ball_L(x)$ by the scatterers supported by $y\not\in \ball^\tau_{L}(x)$ has sup-norm
%%within the stability limits of the projection on $\ball_L(x)$ of any sample
%%$\Bfq' \in \mcA\big(\ball_L(x) \big)$.

\ble[Factorization of the probability of a bad cluster]
\label{lem:factor.prob.S.polynom}
Suppose that
$$
\pr{\ball_{L_k}(u) \text{ is not $(E,m)$-SNS} } \le p_k .
$$
Let $S_{k+1}$ be the maximal cardinality of a collection of
balls $\ball_{L_k}(u_i)$, $i=1, 2, \ldots$, with pairwise $2L_k^{\tau}$-distant admissible centers,
of which neither is $(E,m)$-SNS.
Then for any integer $S\ge 0$
$$
\prob{ S_{k+1} > S} \le  C_d Y_{k+1}^{(S+1)d} p_k^{S+1} \,.
$$
\ele

\proof
Using induction on $j\in[1, S]$, it suffices to prove that, with
$$
\mcA_j = \cup_{i=1}^j \ball_{2L_k^{\tau}(u_i)}\,, \;\; j=1, \ldots, S\,,
$$
one has
\be
\label{eq:fact.prob.ind}
\pr{\ball_{L_k}(u_{j+1}) \text{ is not $(E,m)$-SNS} \cond \fF_{ \mcZ \setminus \mcA_j} } \le p_k.
\ee
By Lemma \ref{lem:cond.SNS.polynom}
the event $\myset{\ball_{L_k}(u) \text{ is not $(E,m)$-SNS} }$
is $\fF_{\ball_{L^\tau_k}(u)}$-measurable, so \eqref{eq:fact.prob.ind} holds true.
Hence the claim follows by induction,  since the number of
all collections of $(S+1)$ admissible centers $u_i\in\ball_{L_{k+1}}(u)$, distant or not, is bounded by $C_dY_{k+1}^{(S+1)d}$.
\qedhere

A reader familiar with the paper \cite{GK01} can see that main ingredients of the proof
of the next lemma follow closely the respective arguments from the Germinet--Klein analysis.

\ble[Scaling of probabilities]
\label{lem:ind.prob.polynom}
Assume that
$$
\sup_{u\in\DZ^d} \; \pr{\ball_{L_k}(u) \text{ is not $(E,m)$-SNS} } \le p_k \le L_k^{-b}
$$
Then
$$
\sup_{u\in\DZ^d} \; \pr{\ball_{L_{k+1}}(u) \text{ is not $(E,m)$-SNS} } \le p_{k+1} \le L_{k+1}^{-b}\,.
$$
\ele

\proof
By Lemma \ref{lem:cond.SNS.polynom}, if $\ball_{L_{k+1}}(u)$ is not $(E,m)$-SNS, then either it is not
$(E,\gam_{k+1}$-CNR or it contains a collection of at least $S+1$ balls $\ball_{L_k}(x_i)$
neither of which is $(E,m)$-SNS, with admissible and pairwise $2L_k^{\tau}$-distant centers.

The probability of the former event is assessed with the help of the Wegner-type estimate
from Theorem  \ref{thm:Wegner}, relying
on the disorder in the balls $\ball_{jL_{k}^{\tau}}(u)\subseteq \ball$.
Even the largest among them, $\ball_{L_{k+1}}(u)$, is surrounded by a belt of width $L_{k+1}^{\tau}$
where the random amplitudes are not fixed hence can contribute to the Wegner estimate with
$\eps = \eps_{R_{k+1}}$,  $R_{k+1} = L_{k+1}^{\tau}$,
hence the same is true for all of these balls:  for any $j$, we have
(cf. \eqref{eq:thm.Wegner.eps} and \eqref{eq:cor.Wegner.bound})
\be
\label{eq:prob.SNR.polynom}
\pr{ \ball_{jL_{k}^{\tau}}(u) \text{ is not $(E, \eps_R)$-SNR} } \le
\left( L_{k+1}^{\tau}\right)^{-A + d +\theta}
\ee
where we are free to choose $\theta>0$ as small as we please (the actual correction to the
power $A-d$ is logarithmic). Since $\tau>(b+1)/(A-d)$, we can pick $\theta$ so small that
$(A-d-\theta)\tau>b+1$, hence the RHS of \eqref{eq:prob.SNR.polynom} is bounded by
$\half L_{k+1}^{-b'-1}$ with $b'>b$.

The total number of such
balls is $Y_{k+1} = L_k^{\alpha-1} = L_{k+1}^{1 - \alpha^{-1}}$, with $1 - \alpha^{-1} < 1$,
therefore,
\begin{align}
\notag
\pr{ \ball_{L_{k+1}^{1+\tau}}(u) \text{ is not $(E, \eps_R)$-CNR} }
&
\le \half L_{k+1}^{-(b'+1) + 1} <  \half L_{k+1}^{-b } .
\end{align}
By Lemma \ref{lem:factor.prob.S.polynom}
$$
\bal
\prob{ S_{k+1} >  S } &\le \frac{ Y_{k+1}^{S+1} }{(S+1)!} p_k^{S+1}
 \le \half \,  L_{k}^{-(S+1) b } \le \half L_{k+1}^{-b} \,,
\eal
$$
whence
$$
\pr{  \ball_{L_{k+1}}(u) \text{ is not $(E,m)$-SNS }}
\le \half L_{k+1}^{-b} + \half L_{k+1}^{-b} =  L_{k+1}^{-b} \,.
$$
$\,$
\qedhere

By induction on $k$, we come to the conclusion of the fixed-energy MSA
under a polynomially decaying interaction.

\btm
Suppose that the ILS estimate
$$
\sup_{u\in\DZ^d} \; \pr{  \ball_{L_0}(u) \text{ is not $(E,m)$-SNS } } \le L_0^{-b}
$$
holds for some $L_0$ large enough, uniformly in $E\in I \subset\DR$. Then for all $k\ge 0$
and all $E\in I$
$$
\sup_{u\in\DZ^d} \;
\pr{  \ball_{L_k}(u) \text{ is not $(E,m)$-SNS } } \le L_k^{-b}.
$$
\etm

In turn, the required ILS estimate is established in Section \ref{sec:ILS.strong.disorder}
for strongly disordered systems, and in Section \ref{sec:ILSthin.tail} for an arbitrary nonzero
amplitude of the potential, with the help of the "thin tails" argument. Therefore, the fixed-energy
bound on the decay of Green functions in the balls of radii $L_k$, $k\in\DN$, is proved under either of these conditions.

This concludes the fixed-energy MSA for the long-range Anderson Hamiltonians on $\DZ^d$ with arbitrary nontrivial
probability distribution of the IID amplitudes $\om_\bullet$ and polynomially decaying interaction potential
$\fu(r) = r^{-A}$, $A>d$.

\section{On derivation of spectral and dynamical localization}

Here we follow closely \lcite{C16a}{Section 5}.

\vskip2mm
Proposition \ref{prop:ETV} is an adaptation of a result by
Elgart \etal \cite{ETV10}, and Proposition \ref{prop:EFC.GK01} is essentially a reformulation
of an argument by Germinet and Klein (cf. \lcite{GK01}{proof of Theorem 3.8}) which substantially
simplified the derivation of strong dynamical localization from the energy-interval MSA bounds,
compared to \cite{DS01}.q

Introduce the following notation: given a ball $\ball_L(z)$ and $E\in\DR$,
$$
\bal
\BF_{z,L}(E) &:= \big| \ball_L(z) \big| \; \max_{|y-z|} \big| G_{\ball_L(z)}(z,y;E) \big|\,,
\eal
$$
with the convention that $\big| G_{\ball_L(z)}(z,y;E) \big| = +\infty$ if
$E$ is in the spectrum of $H_{\ball_L(z)}$. Further, for a pair of balls
$\ball_L(x), \ball_L(x)$ set
$$
\bal
\BF_{x,y,L}(E) &:= \max\, \big[ \BF_{x,L}(E), \, \BF_{y,L}(E) \big] \,.
\eal
$$

The fixed-energy MSA in an interval $E\in I\subset \DR$ provides probabilistic bounds on
the functional $\BF_{x,L}(E)$ of the operator $H_{\ball_L(x)}(\om)$; as a rule, they are easier
to obtain that those on $\sup_{E\in I} \; \BF_{x,y,L}(E)$ (referred to as energy-interval bounds).
Martinelli and Scoppola \cite{MS85} were apparently the first to notice a relation between the two
kinds of bounds, and used it to prove a.s. absence of a.c. spectrum for Anderson Hamiltonians
obeying suitable fixed-energy bounds on fast decay of their Green functions. Elgart, Tautenhahn and
Veseli\'c \cite{ETV10} improved the Martinelli--Scoppola technique, so that energy-interval bounds
implying spectral and dynamical localization could be derived from the outcome of the fixed-energy MSA.

\bpr[Cf. \lcite{C16a}{Proposition 1}, \cite{ETV10}]
\label{prop:ETV}
Let be given a bounded interval $I\subset\DR$, an integer $L\ge 0$ and disjoint balls $\ball_L(x)$,
$\ball_L(y)$. Assume that for some $\ra_L, \rrq_L\in(0,1]$
$$
\sup_{E\in I} \; \max_{z\in \{x,y\}} \; \pr{\BF_{z,L}(E) > \ra_L} \le \rrq_L.
$$
and for some function $f:\, (0,1] \to \DR_+$,
$$
\all \eps\in(0,1] \quad
\pr{ \dist\Big( \Sigma\big(H_{\ball_L(x)}\big), \, \Sigma\big(H_{\ball_L(x)}\big)\Big) \le \eps}
\le f(\eps).
$$
Then
$$
\pr{\sup_{E\in I} \; \BF_{x,y,L}(E) > \max\big[ \ra_L, \rrq^{1/2}_L \big]}
\le \big| I \big| \,\rrq^{1/4}_L + f\left(2\rrq^{1/4}_L \right).
$$
\epr

For any interval $I\subset \DR$,
denote by $\cscB_1(I)$ the set of  bounded Borel functions $\phi:\, \DR\to\DC$ with
$\supp\, \phi\subset I$ and $\| \phi\|_\infty \le 1$.

The next result is based on the Germinet--Klein techniques.

\bpr[Cf. \lcite{C16a}{Theorem 3}, \cite{GK01}]
\label{prop:EFC.GK01}
Assume that the following bound holds for some $\eps>0$, $\rh_L>0$,
$L\in\DN$ and  a pair of balls $\ball_{L}(x), \ball_{L}(y)$ with $|x-y|\ge 2L+1$:
$$
\;  \pr{\sup_{E\in I} \, \BF_{x,y,L}(E) > \eps} \le \rh_L.
$$
Then for any ball $\ball \supset \big(\ball_{L+1}(x) \cup \ball_{L+1}(y) \big)$
$$
\esm{ \sup_{\phi\in\cscB_1(I)} \big| \lr{ \one_x \,|\, \phi\big(H_\ball\big) \, \one_y \,} }
\le 4\eps + \rh_L \,.
$$
\epr

Now the assertions of Theorems \ref{thm:main.polynom.large.g} and \ref{thm:main.polynom.low.E}
follow from the fixed-energy localization analysis carried out in Section \ref{sec:loc.polynom}
with the help of Propositions \ref{prop:ETV} and \ref{prop:EFC.GK01}.

\section{Concluding remarks}

\subsection{What's beyond the piecewise-constant, toy models of interaction?}

Evidently, the analytic form \eqref{eq:def.fu} of $\fu$ is by far too artificial to be physically realistic,
so the question is, to what extent an approximation of a slowly decaying function $\fu:\, r\mapsto r^{-A}$
(i.e., with the derivative $\fu'$ decaying faster than $\fu$ itself) by larger and larger plateaus
on $[\fr_k, \fr_{k+1})$ preserves the regularity properties:
\begin{enumerate}[(i)]
  \item of the single-site distributions of $V(x,\cdot)$ subject to an infinite thermal bath,

  \item of the IDS/DoS in finite balls, again in an infinite thermal bath,

  \item of the finite-ball EV concentration in a "frozen bath".
\end{enumerate}

Of course, in (iii), it is out of question to get any universal \emph{continuity} property for an arbitrary
nontrivial marginal measure of the random amplitudes $\om_\bullet$.
For example, in the Bernoulli case, having at our disposal
only a finite configuration $\myset{\om_x, \, x\in \ball_{L+R}(u)}$, $L,R< \infty$, we get a finite number
of samples of the EVs, hence no bona fide continuity of the IDS. On the other hand, one only needs here
a descent EVC bound, not continuity of the EV distribution.

Curiously, when we replace a random constant potential on a ball $\ball_L(u)$ by an \emph{almost} constant
function $x \mapsto \om_y \fu(|x-y|)$, the resulting effect on the variation of the EVs, which are therefore
no longer \emph{linear} functionals of the r.v. $\om_y$ (\emph{not exactly}, anyway), is akin to introducing
a weak dependence between the variations of any given EV $\lam_j(\om)$ due to distinct amplitudes, say,
$\om_y$ and $\om_z$. This is a subtle effect on which I plan to elaborate in a forthcoming work in the present
series devoted to long-range interactions, but informally it can be explained as follows.

Fix an EV $\lam_j(\om)$ of $H_{\ball_L(u)}(\om)$, and two amplitudes $\om_y, \om_z$ with
$y \ne z$ far away from $\ball_L(u)$. For notational brevity, let $\om_\bullet$ take values $0$ and $1$ (Bernoulli).
Denote $\lam_j(a,b)$ the value of $\lam_j$ with $\om_x=a$, $\om_y=b$, $a,b\in\{0,1\}$. If the interaction,
modulated by $\om_x$ and $\om_y$, were constant $(=C \ne 0)$ on $\ball_L(u)$, we would have
$$
\lam_j(1,0) - \lam_j(0,0) = \lam_j(1,1) - \lam_j(0,1) = C \om_x\,,
$$
which is thus independent of the r.v. $\om_y$. This is precisely the comfortable setting in which we have been
working in the present paper. However, if the interaction is not constant on $\ball_L(u)$, the variations
$$
\lam_j(1,0) - \lam_j(0,0) \;\; \text{ and }\;\; \lam_j(1,1) - \lam_j(0,1)
$$
are no longer identical, generally speaking, hence we have effectively a \emph{dependence} between the variations
of the EVs induced by \emph{independent} random amplitudes $\om_x$, $\om_y$.

Naturally, by the min-max principle, the nonlinear effects (hence, stochastic dependencies) are as weak
as the sup-norm deviation of the potential $\ball_L(u) \ni z \mapsto \fu(|z-x|)$ from an approximating constant
as $z$ runs through the ball $\ball_L(u)$.

The bottom line is that the simplified analysis carried out in this work should be extended in fairly similar
ways
\begin{enumerate}[$\bullet$ ]
  \item to the random media with weak stochastic dependence between the source potentials
  $x \mapsto \om_y \fu(|x-y|)$, and
  \item to the slowly decaying interaction potentials $\fu$, including $r\mapsto r^{-A}$.
\end{enumerate}
In particular, the Gaussian asymptotics should remain valid in a large class of correlated random fields
$\myset{\om_x, \, x\in\DZ^d}$. Different approaches have been developed in the probability theory to
prove limit theorem for sums of weakly dependent r.v.; historically, the first fairly general method was
proposed by S. N. Bernstein \cite{Bern27} in 1927. It has the advantage to be quite straightforward
and flexible. Pictorially, the key idea is that the asymptotic probability distribution for
sequences of random variables $X_n$ et $Y_n$ is the same, provided that
$X_n - Y_n$ has variance relatively small as compared
to $\sigma_n^2 = \esm{ (X_n - \esm{X_n})^2}$. Bernstein's method
applied both to the analysis of the limiting measures
and to their Fourier transforms relies on the linear/bilinear/quadratic functionals of the measures
in question, which makes the analysis explicit and computational.
Certainly, this program requires a number of accurate perturbation estimates.

\subsection{Bernstein-type approximation of the characteristic functions}

We start with auxiliary statements relative to possibly dependent r.v. $X_1, \ldots, X_n$
satisfying the following hypotheses which will be referred to but not repeated every time again:
\be
\label{eq:cond.wek.dep}
\bal
\esm{X_k} &= 0, \;\; m_k = \esm{|X_k|^3} < +\infty\,, \;\; \esm{ X_k^2} =: \sigma_k^2,
\\
S_n &= \sum_{k=1}^n X_k\,, \;\; B_n := \esm{ S_n^2 }, \;_; B'_n := \sum_{k=1}^n \sigma_k^2
\eal
\ee

\be
\bal
 \max_k \;\; \essup \; \big| \esm{X_k \cond \fF_{<k} } - \esm{X_k} \big|
   & \equiv \max_k \;\; \essup \; \big| \esm{X_k \cond \fF_{<k} } \le \alpha_k,
\\
 \max_k \;\; \essup \; \big| \esm{X_k^2 \cond \fF_{<k} } - \esm{X_k^2} \big|
   & \le \beta_k \,,
\\
 \max_k \;\; \esm{X_k^3 \cond \fF_{<k} } & \le c_k \,,
\eal
\ee

Here $\essup(\ldots)$ refers to the dependence of the respective conditional expectations
(which are, by definition, random variables) upon the conditions.

\ble[Cf. \lcite{Bern27}{Section 9, \, Eqns. (65)--(66)}]
\label{lem:Bern27.esm}
Consider $X_k, k=1, \ldots, n$ satisfying \eqref{eq:cond.wek.dep}.

Let $Y_k := X_k/\tB_n^{1/2}$. For any $T<\infty$ there exists a constant
$C_T$ determined solely by $T$ such that for all $|t|\le T$,
$$
\bal
\esm{ \eu^{\ii t Y_k}} &= 1 - \frac{\esm{ X_k^2}}{\tB_n}t^2 + \delta_k(t)\,,
\\
 |\delta_k(t)| & \le C_T\left( \alpha_k \, B_n^{-1/2} + \beta_k\, B_n^{-1} + c_k\, B_n^{-3/2}  \right) =: \eta_k(T) \,.
\eal
$$
\ele

The next result is an adaptation of \lcite{Bern27}{Lemme Fondamental, p.21}, where we focus
on the characteristic functions rather on the respective probability measures.

\ble
Consider $X_k, k=1, \ldots, n$ satisfying \eqref{eq:cond.wek.dep}, let $S_n = X_1 + \cdots + X_n$,
and assume that
\be
 B_n^{-1/2}\sum_{k=1}^n \alpha_k  + B_n^{-1} \sum_{k=1}^n \beta_k + B_n^{-3/2} \sum_{k=1}^n c_k
   \tto{n\to\infty} 0.
\ee
Denote, for $k=1, \ldots, n$,
\be
\ffi_k(t) = \esm{ \eu^{\ii t \sum_{j=1}^k Y_k} } \,, \quad
\Psi_k(t) := \prod_{j=1}^k \left( 1 - \sigma_j^2\, \tB^{-1}_n\,  t^2 \right) \,.
\ee
Then
\be
\sup_{|t|\le T} \; \big| \ffi_n(t) - \Psi_n(t) \big| \le \sum_{k=1}^n \eta_k(T) \,.
\ee

\ele

\proof
Use Lemma \ref{lem:Bern27.esm}. For any pair of r.v., $X$ and $Z$, such that
$$
\esm{ Z \cond X } = \mcR + \delta(X),
$$
for some non-random $\mcR\in\DR$,
$$
\esm{ X Z} = \mcR \esm{ X Z}  + \esm{ X  \, \delta(X) },
$$
thus if $|X|\le C$ and $|\delta(X)|< \eps$, then
$$
\big| \esm{XZ} - \mcR \esm{ X} \big| < C\eps.
$$
Hence
$$
\bal
 \ffi_m = \esm{ \eu^{\ii t \sum_{k=1}^{m-1} Y_k} \cdot \eu^{\ii t Y_m } }
&= \ffi_{m-1} \cdot \left( 1 - \tB_n^{-1/2} \esm{X_m^2}  t^2 \right) + \gamma_m \,,
\;\; \text{ with } \; |\gamma_m| \le \eta_m \,.
\eal
$$
By recurrence,
$$
\bal
\ffi_m &= \Psi_m + \Psi_m \sum_{k=1}^{m} \frac{\gamma_k}{E_k}
\\
\Psi_k &:= \prod_{k=1}^m \left( 1 - \frac{ \esm{X_k^2}}{\tB_n} t^2 \right) \,.
\eal
$$
On note that $\big| \Psi_m/\Psi_k\big| \le 1$, and $\gamma_k \to 0$ as $k\to\infty$. So
$$
\big| \ffi_m - \Psi_m \big| < \sum_{k=1}^m |\gamma_k| \tto{m \to\infty} 0.
$$
This proves the claim; note that in the original statement from \cite{Bern27}, one derives from the above
estimates a perturbation bound on the respective probability distribution and concludes that it
is asymptotically Gaussian (Bernstein focuses in \cite{Bern27} on the proof of a CLT for weakly dependent r.v.).
\qedhere

Now it becomes clear that, in view of the discussion in the previous subsection,
the artificial, piecewise-constant interaction potentials $\fu$ can be effectively used as
convenient approximants for more realistic ones, and such an approximation preserves
the qualitative (as well as some quantitative) estimates resulting in infinite derivability
of the single-site distributions and of the finite-volume DoS. There would be basically no point
in carrying out this program exclusively for the IID random fields $\myset{\om_x, \, x\in\DZ^d}$,
as Bernstein's techniques had been specifically designed to address the (weakly) dependent systems,
and I plan to do so in a separate text.

\subsection{Extremely slow decay of interaction: making use of embedded dipoles}

As was already mentioned in \cite{C16e}, an infinite range of interaction is after all a double-edged sword,
and the regularization effects beneficial to the EVC analysis come with a price of infinite-range stochastic
correlations, particularly inconvenient in the case of weakly screened, barely summable interaction potentials.
A spectacular example is provided by 1D systems where one can have an asymptotic behaviour
(cf., e.g., \cite{PetAntNil91})
$\fu(r) \sim \frac{1}{r\, \ln^2 r}$ for $r \gg 1$. In fact, we have seen that the problem with very slowly
decaying interactions is two-fold: not only the analysis of the effects of correlations becomes more tedious,
but also the "frozen bath" EVC estimates, relying only on disorder in a ball of limited size,
become weaker. One can wonder, therefore, if there is a way to circumvent these difficulties, particularly
the latter one.

Indeed, the situation is not hopeless, as one can benefit from an effect which can be pictorially
described as "screening within screening". We will now consider the case of a realistic potential $\fu$
not approximated by a piecewise constant function, e.g., $\fu(r) = r^{-A}$ or $r^{-1} \ln^{-A} r$, $A>0$,
and Bernoulli amplitudes $\om_\bullet = \pm 1$. If for a pair of neighboring sites $x,y$ one has
$\om_x = -\om_y$, then this pair generates a dipole type potential
$$
z \mapsto \pm\big( \fu(|z-x|) - \fu(|z-y|)  \big)
$$
which behaves at large distances ($|x-z| \gg |x-y|$) like the derivative of $\fu$ in the direction of
the vector $x-y$, hence decays slower than the $\fu(|x-z|)$. Observe that $\fu$ is already a screened
potential, and restricting ourselves to the particular case of a dipole pair, we acquire in the derivative
an additional power $r^{-1}$ in the decay, compared to $r^{-A}$ or, respectively, $r^{-1} \ln^{-A} r$.

Naturally, an occurrence of a dipole pair is a random event, but working for simplicity on the Bernoulli model
and partitioning non-randomly a set of sites $\Lam$ of large cardinality into nearest-neighbor pairs
$(x_i, y_i)$, $i=1 \ldots, |\Lam|/2$, we shall have by the usual CLT roughly half of the pars in the dipole
configurations, and among these dipoles, roughly a half will be of the type $(1, -1)$ and half of the type
$(-1,1)$ -- again by the CLT. Moreover, the standard Large Deviations Estimates (LDE) theory provides
exponentially strong probabilistic bounds on the events where the deviation of the number of desired
configurations from their expected value is $\ge c |\Lam|$, with an arbitrarily small $c>0$.
Thus unwanted, non-typical situations can be ruled out with probabilistic precision exceeding by far
what we need in the analysis of regularity of the DoS.

We conclude that the efficiency of the regularity analysis can be improved
by considering an embedded random dipole model, where the lattice is partitioned from the beginning
into dipoles and non-dipole pairs. Thanks to the LDE, we can safely assume
the quantity of dipoles suffices to repeat our regularity analysis relying only on dipoles
and conditioning on the remaining pairs $(x_i, y_i)$; by doing so, we effectively replace the
potential $\fu$ by its faster-decaying derivatives.

Naturally, one can consider a random dipole (rather than random charge) problem from the beginning.
Technically speaking, it is quite close to a random displacements problem, where not the amplitude $\om_x$
but the position of each charge is random and takes at least two distinct values; in this case the
respective amplitude $\om_x$ can be fixed, as the main contribution to the regularizing convolution
mechanism would come from randomization of the position.

%%

%\bibliography{mybib}   % name your BibTeX data base
%%%%%% Non-BibTeX users please use
%%%%%\begin{thebibliography}{}
%%%%%%
%%%%%\end{thebibliography}

%%%\begin{thebibliography}{}

\end{document}